\documentclass{article}

\PassOptionsToPackage{table}{xcolor}
\usepackage{iclr2027_conference}

\usepackage{iftex}
\ifPDFTeX
  \usepackage{times}
\else
  \usepackage{fontspec}
\fi

\usepackage{amsmath,amssymb,mathtools}
\usepackage{booktabs}
\usepackage{graphicx}
\usepackage{multirow}
\usepackage{microtype}
\usepackage{xcolor}
\usepackage{enumitem}
\usepackage{float}
\usepackage{placeins}
\usepackage{wrapfig}
\usepackage{xspace}
\usepackage{url}
\usepackage{hyperref}

\hypersetup{
  colorlinks,
  citecolor=[rgb]{0.10,0.50,0.15},
  linkcolor=[rgb]{0.75,0.20,0.15},
  urlcolor=[rgb]{0.80,0.15,0.55}
}

\definecolor{methodgreen}{RGB}{18,133,95}
\definecolor{resultblue}{RGB}{42,120,214}
\definecolor{stdgray}{gray}{0.45}
\definecolor{cBest}{RGB}{231,243,238}
\definecolor{cSecond}{RGB}{238,241,246}
\definecolor{cPos}{RGB}{58,164,110}
\definecolor{cNeg}{RGB}{217,106,89}

\newcommand{\method}{\textsc{EmoUpdate}\xspace}

\newcommand{\sba}{S-BAcc\xspace}
\newcommand{\tfone}{Transition F1\xspace}
\newcommand{\best}[1]{\ifmmode\mathbf{#1}\else\textbf{#1}\fi}
\newcommand{\ours}[1]{\textcolor{methodgreen}{\textbf{#1}}}
\newcommand{\stdv}[1]{{\color{stdgray}\tiny$\pm#1$}}

\title{
Do SpeechLMs Hear Their Own Opinions?\\
Diagnosing and Mitigating Previous-Belief\\
Contamination in Streaming Emotion\\
Understanding
}

\author{
\textbf{Haoyue Liu}\textsuperscript{1}
\quad
\textbf{Zhichao Wang}\textsuperscript{1}
\quad
\textbf{Ye Chen}\textsuperscript{2}
\quad
\textbf{Haonan Deng}\textsuperscript{4}
\quad
\textbf{Xiaoying Tang}\textsuperscript{1,3,\ensuremath{\dagger}}
\\[0.6em]
\textsuperscript{1}
School of Science and Engineering,
The Chinese University of Hong Kong, Shenzhen 518172, China
\\
\textsuperscript{2}
XJTU-POLIMI Joint School,
Xi'an Jiaotong University, Xi'an 710049, China
\\
\textsuperscript{3}
Shenzhen Future Network of Intelligence Institute (FNii-Shenzhen)
\\
\textsuperscript{4}
University of California, Berkeley
}

\iclrfinalcopy

\begin{document}

\maketitle

\begin{abstract}
Streaming emotion understanding uses historical state while continuously interpreting current audio, often feeding the model's previous prediction back as context. We show that this history conditioning can distort current perception. On a balanced CREMA-D-Stream counterfactual diagnostic, changing only the injected previous emotion label while holding the audio fixed reduces current-audio accuracy from 72.50\% to 30.42\% and flips 65.69\% of predictions. The effect is strongly label-asymmetric, with prior pull ranging from 4.76\% to 98.20\%, revealing a failure we call \emph{previous-belief contamination} (PBC). To address PBC, we introduce \method, a training-free framework that separates current-audio perception from historical state revision through three components: \textbf{(1)} a \emph{prior-blind acoustic firewall} that prevents historical state from entering perception; \textbf{(2)} an \emph{evidence-shrunk causal belief filter} that introduces history only after observation formation and retains label-asymmetric transition structure only when supported by observed evidence; and \textbf{(3)} a closed-form \emph{decontamination operator} derived from the same counterfactual measurements for serving stacks where firewalling is unavailable.
Across four SpeechLMs and two streaming emotion benchmarks, \method achieves the best step accuracy and state-balanced accuracy in all eight model--benchmark settings, improving S-BAcc by up to 69.71 points and step accuracy by up to 38.41 points over the strongest controlled baselines.
\end{abstract}

% ---------- sections/introduction.tex ----------
\section{Introduction}
\label{sec:introduction}

Audio-language models can reason directly over speech without first reducing it to a transcript~\citep{chu2024qwen2,xu2025qwen25omnitechnicalreport,abouelenin2025phi}, making them well suited to \emph{streaming} emotion understanding, where a system tracks a speaker's emotional state as audio continuously arrives~\citep{guo2026emos,wang2026humdial,song2026learning}. Such tasks require a model to interpret the current audio while also using historical state, so its previous prediction is fed back as historical context. We formalize this design below as \emph{direct history conditioning}.

We ask how this history conditioning affects current perception. Once the previous state enters the perception prompt, it is no longer only a summary of the past. It becomes textual context that directly interacts with the current acoustic evidence. The model may copy the previous judgment, preserve it through a genuine emotional shift, or more subtly reshape the current posterior. The system, in short, begins to hear its own opinions. This failure can be mistaken for ordinary recognition error, but it is structurally different. The current audio remains unchanged, yet the resulting observation changes when only the injected historical label is replaced. It also differs from context-induced biases caused by external textual evidence~\citep{goyal2024context}. Here, the contaminating variable is \emph{self-generated and recurrent}, so an erroneous state can re-enter subsequent inference and propagate over time.

\begin{figure}[t]
    \centering
    \includegraphics[width=\textwidth]{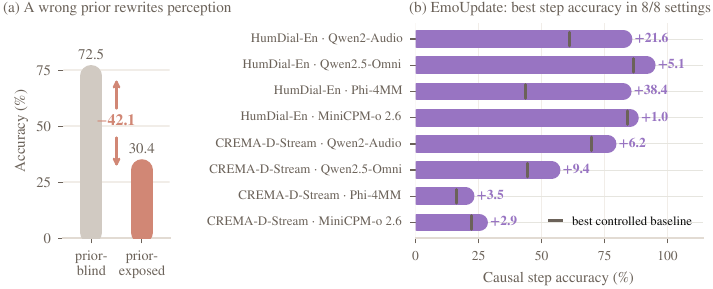}
    \caption{\textbf{Previous-belief contamination and its repair.}
    (a)~A wrong injected prior collapses current-audio accuracy (CREMA-D-Stream,
    Qwen2-Audio). (b)~\method attains the best causal step accuracy in all
    eight combinations of models and benchmarks.}
    \label{fig:hero}
\end{figure}

To make this failure measurable, we introduce a counterfactual intervention protocol that holds the current audio, instruction, and decoding fixed while varying only the injected previous emotion label and measuring the resulting posterior change. On a balanced CREMA-D-Stream diagnostic, this intervention reveals a failure we call \emph{previous-belief contamination} (PBC): history exposure reduces current-audio accuracy by 42.08 points, flips 65.69\% of predictions, and pulls 71.00\% of wrong-prior conditions toward the injected label. More importantly, the effect is strongly label-asymmetric, with prior pull ranging from 4.76\% to 98.20\% across emotion labels. A single uniform stickiness or global calibration parameter is therefore insufficient to characterize the contamination. Prompt wording alone is also insufficient: seven of ten grounding candidates explicitly instruct the model to ignore the previous label, yet wrong-prior pull never falls below 59.7\%. These results suggest that the problem is not simply how strongly history is weighted. Historical state participates in forming the current observation itself, which explains both the degradation under direct history conditioning and the gains from separating perception from state revision shown in Figure~\ref{fig:hero}.

We therefore introduce \method (Figure~\ref{fig:overview}), a training-free streaming inference framework. Its key idea is not to further adjust the weight of historical information, but to change where history enters the system. The current audio first forms an independent acoustic observation, and historical state is introduced only afterward for revision. 
Specifically, \method \textbf{(1)} isolates historical state during perception so that current acoustic evidence is not directly shaped by the previous prediction; \textbf{(2)} performs post-perception causal revision with an evidence-shrunk transition rule that preserves label-asymmetric dynamics only when supported by transition evidence and otherwise reduces to a symmetric prior; and \textbf{(3)} converts the same counterfactual measurements into a closed-form decontamination repair when such isolation is unavailable at the serving interface. Historical information therefore no longer participates in extracting current acoustic evidence and instead acts only during subsequent state revision.

We summarize our contributions as follows:
\begin{itemize}[leftmargin=14pt,nosep]
    \item \textbf{We identify and quantify previous-belief contamination (PBC), a self-referential failure mode in streaming SpeechLMs.} Controlled interventions show that a model's own previous prediction can alter its perception of identical audio, with strong label asymmetry.

    \item \textbf{We introduce \method, a streaming inference paradigm that separates current perception from historical state revision.} It first forms a prior-blind acoustic observation, then performs evidence-shrunk causal revision that retains label-asymmetric transitions only when supported by evidence. When firewalling is unavailable, the same measurements yield a closed-form decontamination repair without training or extra deployment calls.

    \item \textbf{We systematically validate the effectiveness and mechanism of \method.} Across four SpeechLMs and two streaming emotion benchmarks, \method achieves the best step accuracy and state-balanced accuracy in all eight settings, with gains of up to 38.41 and 69.71 percentage points over the strongest controlled baselines.
\end{itemize}

% ---------- sections/related_work.tex ----------
\section{Related Work}
\label{sec:related}

\paragraph{Speech emotion understanding and streaming emotion modeling.}
Large audio-language models increasingly support direct speech input and instruction following~\citep{chu2024qwen2,xu2025qwen25omnitechnicalreport,abouelenin2025phi,tang2024salmonn,gong2024listen}. Recent work studies speech emotion through prosodic prompting and lexical--paralinguistic conflict~\citep{wang2026vowelprompt,pang2026audio}, while HumDial-EIBench and EmoS evaluate multi-turn and streaming emotional trajectories~\citep{wang2026humdial,guo2026emos}. More broadly, conversational emotion recognition models temporal dependence through emotional inertia, shift-aware context, speaker-conditioned priors, and learned update controllers~\citep{liu2024emotionic,zha2025dual,kaplan2026scope,shen2024emotion,song2026learning}. These approaches primarily model how emotion evolves over time. We study a different question: whether the system's own previous prediction changes its perception of identical current speech. We diagnose this effect through direct intervention on the previous state and separate current-audio perception from subsequent historical state revision.

\paragraph{Self-conditioning, prompt bias, and error accumulation.}
Conditioning a model on its own previous outputs is a classical source of error accumulation~\citep{bengio2015scheduled,ranzato2015sequence,lamb2016professor,arora2022exposure}, while prompted language models can also be influenced by asserted beliefs, prompt position, or conflicting context~\citep{sharma2024towards,xu2024earth,huang2024large,zhao2021calibrate,liu2024lost,goyal2024context}. We focus on a different closed-loop setting in which the model's own previous output re-enters the next perception prompt. PBC differs from ordinary exposure bias because the previous self-generated label does not merely affect a later prediction; it changes the model's observation of unchanged current audio itself. We expose this failure through controlled counterfactual intervention and separate current-audio grounding from subsequent historical state revision. Appendix~\ref{app:related} expands this comparison and further covers speech representations and emotion corpora.

% ---------- sections/method.tex ----------
\section{Method}
\label{sec:method}

\begin{figure}[t]
    \centering
    \includegraphics[
        width=\textwidth,
        height=0.5\textwidth,
        keepaspectratio
    ]{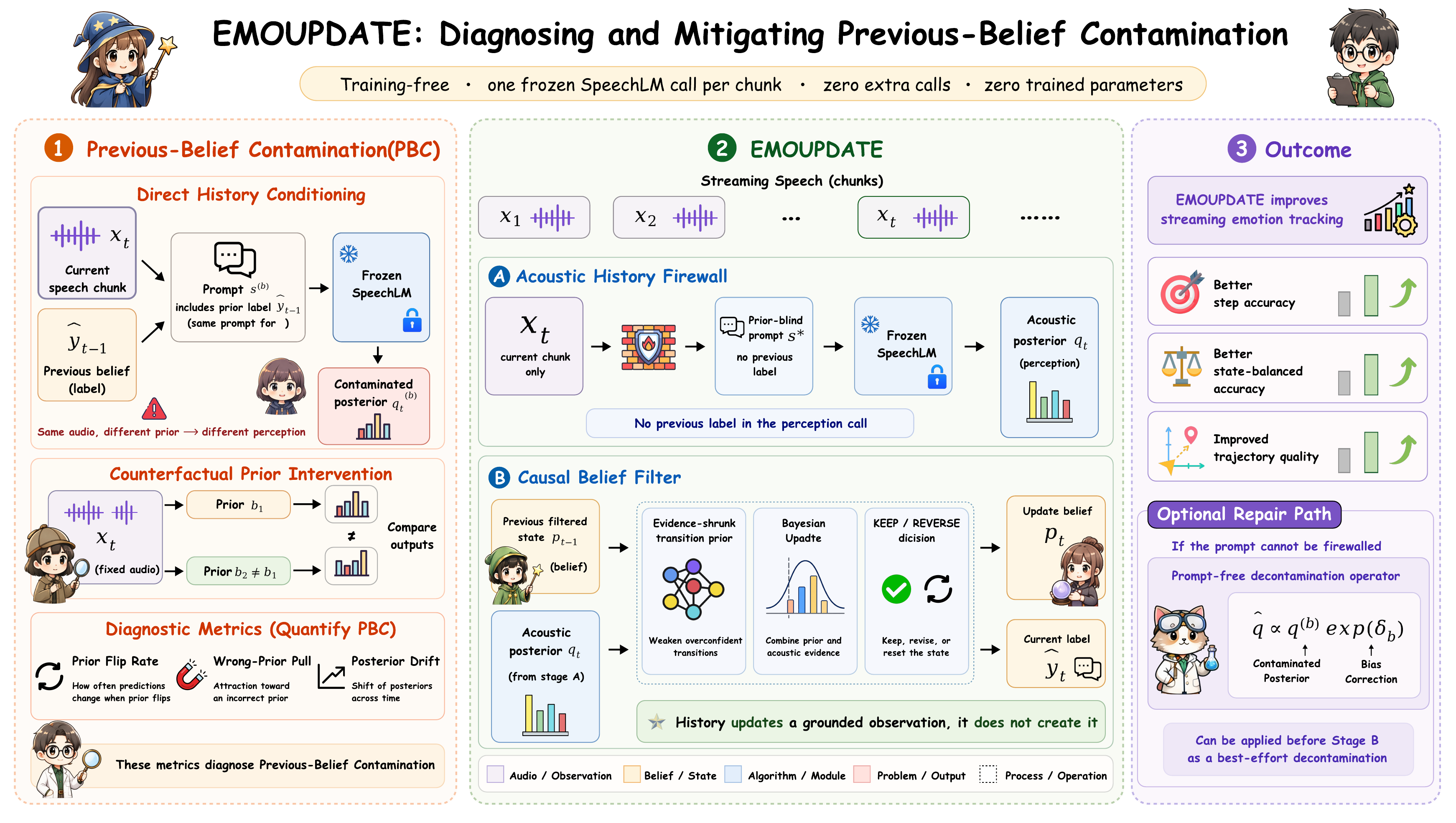}
    \caption{\textbf{From entangled perception to causal belief revision.}
    (1)~Direct history conditioning writes the previous label into the
    perception prompt; a counterfactual intervention on that label quantifies
    previous-belief contamination. (2)~\method restores the observation with a
    prior-blind acoustic firewall, then revises it with an evidence-shrunk
    causal filter. (3)~Trajectory quality improves; a prompt-free
    decontamination operator is the fallback where firewalling is unavailable.}
    \label{fig:overview}
\end{figure}

We formulate streaming emotion understanding as causal state estimation, then
describe the intervention used to diagnose PBC and the three components of
\method: the acoustic history firewall, the evidence-shrunk causal belief
filter, and the decontamination fallback for stacks that cannot be firewalled
(Figure~\ref{fig:overview}).

\subsection{Streaming Emotion Belief Revision}

An episode is a sequence $\mathcal{E}=\{(x_t,y_t)\}_{t=1}^{T}$ of $T$ chunks,
where $x_t$ is the current speech chunk, $\mathcal{Y}$ is the fixed emotion
ontology, and $y_t\in\mathcal{Y}$ is the state of that chunk. After
receiving $x_t$, a causal system outputs a belief distribution $p_t$ and label
$\hat y_t=\arg\max_y p_t(y)$ without observing future audio. The update action
is \textsc{keep} when $\hat y_t=\hat y_{t-1}$ and \textsc{revise} otherwise.
This exposes two opposing errors: \emph{inertia}, retaining a previously correct
label after a true shift, and \emph{volatility}, destroying a correct belief
while the gold state remains stable.

A direct history-conditioned SpeechLM computes
\begin{equation}
q_t^{(b)} = f_{\theta}(x_t, s, b_{t-1}),
\label{eq:entangled}
\end{equation}
where $q_t^{(b)}$ is the resulting label posterior over $\mathcal{Y}$,
$f_{\theta}$ is the frozen SpeechLM, $s$ is an instruction, and
$b_{t-1}\in\mathcal{Y}$ is the previous emitted state label, rendered into the
prompt as text. In deployment, $b_{t-1}=\hat y_{t-1}$.
Equation~\ref{eq:entangled} therefore uses $b_{t-1}$ both as temporal evidence
and as text inside the mechanism that extracts current evidence. Our central
claim is that these roles must be separated.

\subsection{Counterfactual Previous-Belief Intervention}
\label{sec:intervention}

For a fixed current chunk $x_t$, we first obtain the control posterior
$q_t^{(\varnothing)}=f_{\theta}(x_t,s)$, then insert each candidate previous
label $b\in\mathcal{B}\subseteq\mathcal{Y}$ while keeping all other inputs fixed:
\begin{equation}
q_t^{(b)}=f_{\theta}(x_t,s,b), \qquad b\in\mathcal{B}.
\end{equation}
We quantify PBC with three complementary measures. \emph{Prior flip rate}
counts interventions for which
$\arg\max q_t^{(b)}\neq\arg\max q_t^{(\varnothing)}$. \emph{Wrong-prior pull}
is the fraction of conditions with $b\neq y_t$ for which
$\arg\max q_t^{(b)}=b$. Finally, posterior drift averages the Jensen--Shannon
divergence $D_{\mathrm{JS}}(q_t^{(b)}\|q_t^{(\varnothing)})$. The test is
diagnostic: its multiple calls are never used during deployment.

\subsection{Acoustic History Firewall}
\label{sec:firewall}

\method constrains the perception input: the call may not receive the previous
label, filtered state, episode history, future audio, speaker identifier, or
filename cue. Each call is stateless: no conversation history, no carried
system prompt, and no reused decoding context. Every prompt actually sent is
recorded by SHA-256 in the released audit logs, so prior-blindness is checkable
rather than asserted. It computes
\begin{equation}
q_t=f_{\theta}(x_t,s^{\star}),
\label{eq:priorblind}
\end{equation}
using only the current audio and the selected acoustic instruction $s^{\star}$,
which asks for pitch, energy, rhythm, pace, vocal tension, and voice quality,
and for exactly one ontology label. The posterior is read densely from
label-token probabilities, not from self-reported confidence.

The firewall is an input constraint, not a particular prompt string. We
instantiate it with a fixed ten-candidate development pool whose members use
compact alternative strategies for eliciting current-audio evidence while
preserving the same prior-blind input contract. Selection uses development data
only and is locked before evaluation under a pre-specified trajectory-metric
ranking with a final-accuracy floor (Appendix~\ref{app:prompts} states the exact
rule and lists every candidate). This is black-box selection over readable
instructions, not SFT, RL, or soft-prompt learning.

\subsection{Causal Belief Filter}
\label{sec:filter}

The prior-blind posterior is fused with history only after perception. Here
\emph{causal filtering} denotes strictly forward online inference using only
current and past observations; the effect of prior exposure is established
separately by the controlled intervention of Section~\ref{sec:intervention}.
Let $\rho\in[0,1]$ denote the probability that the emotion remains unchanged and
$K=|\mathcal{Y}|$. The one-step predictive prior is
\begin{equation}
\bar p_t(y)=\rho p_{t-1}(y)+\frac{1-\rho}{K-1}\left(1-p_{t-1}(y)\right).
\end{equation}
We then apply a causal Bayesian update,
\begin{equation}
p_t(y)=\frac{q_t(y)\bar p_t(y)}
{\sum_{y'\in\mathcal{Y}}q_t(y')\bar p_t(y')}.
\label{eq:filter}
\end{equation}
This is the filtering recursion of a symmetric first-order hidden Markov
model~\citep{rabiner1989tutorial}, with $p_1=q_1$ and $\rho$ selected from five
pre-specified development values; the filtered argmax induces an explicit
\textsc{keep}/\textsc{revise} action. This symmetric filter is the base case
extended by the evidence-shrunk transition prior in the next subsection. The
decontamination operator instead provides a separate fallback when the
perception interface cannot be firewalled.

\subsection{Evidence-Shrunk Transition Prior}
\label{sec:shrinkage}

The uniform off-diagonal term treats all revisions as exchangeable. We
therefore let development transition statistics encode label-asymmetric
destinations, shrunk toward the uniform null when evidence is weak. For emotion
states $i,j\in\mathcal{Y}$, the off-diagonal transition is
$A(i{\to}j)=(1-\rho)L_\alpha(j\,|\,i)$, where $A$ denotes the transition kernel
as in~\citet{rabiner1989tutorial}. Here $L_\alpha$ shrinks the add-one-smoothed
transition estimate $L$ toward the uniform null using the positive-part form
of~\citet{james1961estimation}:
\begin{equation}
L_\alpha(j\,|\,i)=\tfrac{1-\alpha}{K-1}+\alpha L(j\,|\,i),
\qquad
\alpha=\left(1-\tfrac{\mathrm{df}}{\chi^2}\right)^{+},
\label{eq:shrinkage}
\end{equation}
where $(z)^+=\max(z,0)$, $\chi^2$ is Pearson's statistic for the development
destination counts against the uniform null, and $\mathrm{df}$ is its degrees
of freedom. Since $\mathbb{E}[\chi^2]=\mathrm{df}$ under the null, $\alpha=0$
whenever the observed statistic does not exceed its null expectation, causing
the filter to reduce to Equation~\ref{eq:filter} rather than fitting a spurious
transition table. We use this positive-part form as a closed-form shrinkage rule
toward the uniform null; Appendix~\ref{app:asym} reports its bootstrap
stability. The estimate is closed-form and adds no tuned scalar, no training,
and no deployment cost.

\subsection{Prompt-Free Decontamination Operator}
\label{sec:decon}

When the perception prompt cannot be firewalled, as in legacy or third-party
serving stacks, the same intervention grid yields a repair rather than a
diagnosis. Writing $q^{(\varnothing)}$ for the clean posterior and $q^{(b)}$ for
the one contaminated by previous label $b$, we estimate a per-label log-offset
and invert it,
\begin{equation}
\delta_b(y)=\mathbb{E}\!\left[\log q^{(\varnothing)}(y)-\log q^{(b)}(y)\right],
\qquad
\hat q \propto q^{(b)}\exp(\delta_b).
\label{eq:decon-main}
\end{equation}
The estimate is closed-form, needs no prompt access, no gradient training, and
no extra SpeechLM calls. It requires only the label that the deployed system
itself wrote into the prompt. We fit $\delta_b$ on one stratified half of the
grid and apply it to the held-out half
(Appendix~\ref{app:decontamination}).

\paragraph{Budget-matched endpoint calibration.}
A frozen SpeechLM's label posterior carries its own class prior, which need not
match the benchmark's. We therefore optionally calibrate the terminal
observation by the development prior ratio
$w(y)=\hat\pi_{\text{gold}}(y)/\hat\pi_{\text{pred}}(y)$, the standard
prior-adjustment correction for label shift~\citep{saerens2002adjusting,
lipton2018detecting}; dividing a discriminative posterior by its class prior to
obtain an HMM-compatible likelihood is the scaled-likelihood construction of
hybrid speech recognition~\citep{bourlard2012connectionist}. This adds no
search budget: five values of $\rho$ are paired with two endpoint modes,
forming ten pre-specified filter configurations. These are separate from the
ten grounding-prompt candidates in Section~\ref{sec:firewall}. Both are selected
using development data only and locked before evaluation, while $\alpha$ and
$w$ are estimated from development labels only.

% ---------- sections/experiments.tex ----------
\section{Experiments}
\label{sec:experiments}

In this section we carry out experiments to address the following questions:
\begin{itemize}[leftmargin=14pt,nosep]
    \item \textbf{Q1}: Does a previous belief change what a frozen SpeechLM
    hears from fixed audio? See \S\ref{answer1}.
    \item \textbf{Q2}: Where does history matter most? See \S\ref{answer1b}.
    \item \textbf{Q3}: Does decoupling improve streaming trajectories? See
    \S\ref{answer2}.
    \item \textbf{Q4}: Which components drive the gain? See \S\ref{answer3}.
    \item \textbf{Q5}: Is the result robust, and can the measurements themselves
    repair the failure? See \S\ref{answer5}.
\end{itemize}
The appendix adds (1) per-metric tables, (2) mechanism diagnostics,
(3) difficulty strata, (4) grid-size and (5) bootstrap studies, (6) prompt and
baseline rules, and (7) additional stress tests.

\subsection{Experimental Setup}
\label{sec:setup}

\paragraph{Benchmarks and models.}
We evaluate on two complementary benchmarks
(Table~\ref{tab:data-audit}, Appendix~\ref{app:data-construction}).
\textbf{CREMA-D-Stream} contains four-chunk same-speaker trajectories
that either remain stable or switch emotion after two chunks, with disjoint
development and evaluation speakers.
\textbf{HumDial-En} converts the English emotion-trajectory task of
HumDial-EIBench~\citep{wang2026humdial} into causal chunks of human-recorded
multi-turn speech, with disjoint source groups.
We freeze Qwen2-Audio-7B-Instruct~\citep{chu2024qwen2},
Qwen2.5-Omni-7B-AWQ~\citep{xu2025qwen25omnitechnicalreport},
Phi-4-multimodal-instruct~\citep{abouelenin2025phi}, and
MiniCPM-o-2.6~\citep{yao2024minicpm}, using temperature-zero decoding.

\paragraph{Baselines and compute.}
We compare with direct history conditioning (DHC), independent chunk inference
(ICI), confidence- and margin-gated revision (CGR/MGR), causal HMM filtering
(C-HMM), and instantaneous/accumulated JSD revision (IJSR/AJSR).
Each tunable baseline policy receives one scalar searched over the same
ten-value development grid (Appendix~\ref{app:baselines}).
All controlled policies replay the same posterior cache with one SpeechLM call
per chunk, zero policy calls, and zero trainable parameters. Grounding-prompt
search evaluates the fixed ten-candidate instruction pool once and is reported
separately (Appendix~\ref{app:compute}).

\paragraph{Metrics and selection.}
Our primary metric is \sba, averaging accuracy over true-change and stable
steps; we additionally report step macro-F1, step accuracy, and final accuracy.
\tfone, inertia, volatility, and revision delay serve as mechanism diagnostics
(Appendix~\ref{app:full-results}). All belief policies and $\rho$ values are
selected on development data using the same pre-specified ranking
(\sba, step macro-F1, \tfone, step accuracy, final accuracy), while grounding
prompts follow Section~\ref{sec:firewall}. All choices are locked before
evaluation. We report full-set point estimates with $B=1000$ episode-bootstrap
standard errors (Appendix~\ref{app:reproducibility}).

\subsection{\label{answer1}A1: Previous Beliefs Contaminate Current Perception}

\begin{wrapfigure}{r}{0.47\textwidth}
\vspace{-1.15\intextsep}
\centering
\includegraphics[width=0.45\textwidth]{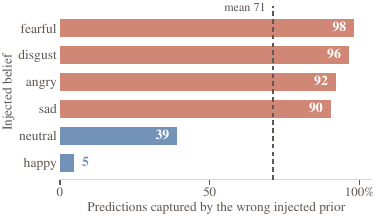}
\caption{\textbf{Counterfactual previous-label intervention}
(CREMA-D-Stream, Qwen2-Audio).
Eligible predictions captured by the wrong injected previous label
(terracotta: injected label wins). A twentyfold spread shows that no uniform
stickiness repairs the effect. Destinations:
Figure~\ref{fig:intervention-dest}.}
\label{fig:intervention}
\vspace{-1.1\intextsep}
\end{wrapfigure}

We sample 120 balanced CREMA-D-Stream development chunks (20 per emotion) and
inject all six candidate previous labels, giving 720 injection conditions
(600 pull-eligible); the asymmetry estimates below are chunk-level, not
episode-level. As Figure~\ref{fig:intervention} shows, changing only the
injected previous label strongly shifts the output: accuracy falls from
72.50\% without previous-label exposure to 30.42\% under exposure, and
65.69\% of predictions flip. Among wrong injected labels, 71.00\% pull the
output to the injected label, while audible evidence overrides one in only
23.68\% of conditions where the control was correct; mean posterior JSD is
0.4823.

The cost is visible at deployment scale on every backbone
(Table~\ref{tab:main-results}; Figure~\ref{fig:contamination-all},
Appendix~\ref{app:intervention}): direct history conditioning loses up to
49 step-accuracy points, and \method is highest in all eight settings.
Instruction-only mitigation remains insufficient in this diagnostic: across
all ten grounding candidates, seven of which explicitly forbid using the
previous label, wrong-prior pull stays above 59.7\% and \emph{rises} with
prior-blind accuracy (Pearson $r=0.81$). This motivates removing historical
belief from the perception call rather than merely instructing the model to
ignore it (Figure~\ref{fig:prompt-frontier},
Appendix~\ref{app:intervention}).

\subsection{\label{answer1b}A2: History Matters Exactly Where Perception Is Weak}

Table~\ref{tab:ambiguity} localizes both the value and the danger of history
across the three backbones on which the balanced injection was run. Call a
chunk \emph{ambiguous} when its prior-blind posterior margin falls below the
development median. There, a \emph{correct} previous label is worth $+17.2$
to $+40.9$ accuracy points on every backbone and benchmark, while a
\emph{wrong} one is adopted in $26.4$--$86.9\%$ of eligible cases. On easy
chunks the policies become numerically indistinguishable because the audio
already settles the label. Benefit and hazard live on the same subset, which
is why history must enter through a filter rather than through the prompt.

Replaying every policy on one shared cache confirms that \method collects this
benefit wherever the acoustic posterior supports it: $+24.07$ for Qwen2-Audio
($p<10^{-4}$) and $+8.66$ for Phi-4MM ($p=0.019$) on HumDial-En, and $+3.07$
on CREMA-D-Stream ($p=0.040$), against at most $+1.28$ on easy chunks in any
setting. The gain is therefore not a uniform prompt effect. When the
prior-blind posterior carries little usable acoustic information, as on
CREMA-D-Stream with Phi-4MM and MiniCPM-o, both below 25\% chunkwise accuracy,
the policy differences become correspondingly small;
Appendix~\ref{app:ambiguity} reports the full stratification.

% ---------- sections/ambiguity_table.tex ----------
% Auto-generated by gen_main_table.py -- do not edit by hand.
\begin{table}[t]
\caption{\textbf{History is worth most exactly where perception is weakest.}
A chunk is \emph{ambiguous} when its prior-blind posterior margin falls below
the development median and \emph{easy} when margin and duration are both above
it, so the two groups are not complementary; thresholds are never tuned on test. $\Delta$correct is the
gain from supplying the \emph{correct} previous label, pull the rate at which
a \emph{wrong} one is adopted, and $\Delta$\method the gain over independent
chunk inference on one shared posterior cache (exact paired McNemar $p$). All
five difficulty groups are in Appendix~\ref{app:ambiguity}.}
\label{tab:ambiguity}
\centering
\scriptsize
\setlength{\tabcolsep}{4.2pt}
\begin{tabular*}{\textwidth}{@{\extracolsep{\fill}}llrrrrrr}
\toprule
 & & \multicolumn{4}{c}{Ambiguous chunks} & \multicolumn{2}{c}{Easy chunks} \\
\cmidrule(lr){3-6}\cmidrule(l){7-8}
Data & Model & $\Delta$correct & Pull & $\Delta$\method & $p$ &
$\Delta$correct & $\Delta$\method \\
\midrule
HumDial-En & Qwen2-Audio & \cellcolor{cPos!35}+39.29 & 63.7 &
\cellcolor{cPos!43}+24.07 & $<$1e--4 & \cellcolor{cPos!8}+0.00 &
\cellcolor{cPos!8}+0.00 \\
 & Phi-4MM & \cellcolor{cPos!25}+27.27 & 30.3 &
\cellcolor{cPos!16}+8.66 & 0.019 & \cellcolor{cPos!8}+5.88 &
\cellcolor{cPos!8}+0.00 \\
 & MiniCPM-o 2.6 & \cellcolor{cPos!16}+17.24 & 26.4 &
\cellcolor{cPos!8}+2.74 & 0.125 & \cellcolor{cPos!8}+0.00 &
\cellcolor{cPos!8}+0.00 \\
\midrule
CREMA-D-Stream & Qwen2-Audio & \cellcolor{cPos!32}+36.07 & 86.9 &
\cellcolor{cPos!8}+3.07 & 0.040 & \cellcolor{cPos!8}+4.35 &
\cellcolor{cPos!8}+0.00 \\
 & Phi-4MM & \cellcolor{cPos!23}+25.86 & 49.0 &
\cellcolor{cPos!8}+0.58 & 0.664 & \cellcolor{cPos!39}+42.86 &
\cellcolor{cPos!8}+0.00 \\
 & MiniCPM-o 2.6 & \cellcolor{cPos!37}+40.91 & 63.6 &
\cellcolor{cPos!8}+0.63 & 0.749 & \cellcolor{cPos!21}+23.53 &
\cellcolor{cPos!8}+1.28 \\
\bottomrule
\end{tabular*}
\end{table}

\subsection{\label{answer2}A3: Decoupling Improves Streaming Trajectories}

Table~\ref{tab:main-results} reports the complete controlled comparison; its
$\Delta$ row is the strict per-metric envelope, so the reference values need
not come from a single baseline.

% ---------- sections/main_results_table.tex ----------
% Auto-generated by gen_main_table.py -- do not edit by hand.
\begin{table}[t]
\caption{Controlled comparison across four frozen SpeechLMs (\%, point estimates
with bootstrap standard errors; shared posterior cache, one call per chunk).
Primary metrics first: state-balanced accuracy and class-balanced step
macro-F1, then step accuracy and conventional final-turn accuracy. Bold:
per-column best; shaded row: \method; $\Delta$: \method minus the per-metric
best controlled baseline.}
\label{tab:main-results}
\centering
\scriptsize
\setlength{\tabcolsep}{2.6pt}\renewcommand{\arraystretch}{0.98}
\resizebox{\textwidth}{!}{%
\begin{tabular}{llrrrrrrrr}
\toprule
 & & \multicolumn{4}{c}{HumDial-En} &
 \multicolumn{4}{c}{CREMA-D-Stream} \\
\cmidrule(lr){3-6}\cmidrule(l){7-10}
Model & Method & S-BAcc $\uparrow$ & Step-MaF1 $\uparrow$ &
Step $\uparrow$ & Final $\uparrow$ & S-BAcc $\uparrow$ &
Step-MaF1 $\uparrow$ & Step $\uparrow$ & Final $\uparrow$ \\
\midrule
Qwen2-Audio & DHC & 1.43\stdv{0.63} & 20.68\stdv{1.81} &
20.00\stdv{1.83} & 4.76\stdv{2.08} & 18.28\stdv{2.18} &
10.97\stdv{1.39} & 20.70\stdv{2.17} & 18.85\stdv{2.50} \\
 & ICI & 29.14\stdv{1.94} & 68.28\stdv{2.00} & 60.95\stdv{2.15} &
\best{89.52}\stdv{3.00} & 70.33\stdv{2.61} & 68.41\stdv{2.01} &
69.57\stdv{2.20} & 70.08\stdv{2.95} \\
 & CGR & 21.71\stdv{1.36} & 55.73\stdv{2.13} & 53.97\stdv{2.27} &
72.38\stdv{4.19} & 70.00\stdv{2.63} & 68.25\stdv{2.02} &
69.57\stdv{2.21} & 70.08\stdv{2.95} \\
 & MGR & 24.57\stdv{1.25} & 60.73\stdv{2.07} & 57.14\stdv{2.17} &
81.90\stdv{3.67} & 70.25\stdv{2.63} & 68.48\stdv{2.01} &
69.88\stdv{2.21} & 70.49\stdv{2.98} \\
 & C-HMM & 29.43\stdv{2.54} & 67.26\stdv{2.08} & 60.00\stdv{2.34} &
85.71\stdv{3.37} & 70.33\stdv{2.61} & 68.66\stdv{2.00} &
69.98\stdv{2.20} & 70.90\stdv{2.97} \\
 & IJSR & 30.00\stdv{2.68} & 67.81\stdv{2.05} & 60.63\stdv{2.25} &
87.62\stdv{3.16} & 69.18\stdv{2.77} & 67.15\stdv{2.27} &
68.55\stdv{2.42} & 67.21\stdv{3.10} \\
 & AJSR & 23.43\stdv{1.31} & 59.82\stdv{2.15} & 55.87\stdv{2.26} &
78.10\stdv{3.90} & 70.66\stdv{2.73} & 68.71\stdv{2.19} &
69.98\stdv{2.34} & 69.67\stdv{3.05} \\
\rowcolor{cBest}
 & \method & \best{70.57}\stdv{4.16} & \best{82.31}\stdv{2.21} &
\best{82.54}\stdv{2.02} & 87.62\stdv{3.27} &
\best{74.02}\stdv{2.58} & \best{74.58}\stdv{1.77} &
\best{76.23}\stdv{2.06} & \best{77.46}\stdv{2.72} \\
 & {\itshape $\Delta$ vs.\ best ctl.} &
\cellcolor{cPos!65}+40.57 & \cellcolor{cPos!22}+14.03 &
\cellcolor{cPos!35}+21.59 & \cellcolor{cNeg!8}$-$1.90 &
\cellcolor{cPos!8}+3.36 & \cellcolor{cPos!9}+5.87 &
\cellcolor{cPos!10}+6.25 & \cellcolor{cPos!10}+6.56 \\
\midrule

Qwen2.5-Omni & DHC & 73.14\stdv{4.28} & 76.63\stdv{2.44} &
76.19\stdv{2.42} & 80.95\stdv{3.81} & 26.15\stdv{2.25} &
26.77\stdv{1.43} & 30.53\stdv{1.83} & 24.18\stdv{2.70} \\
 & ICI & 81.43\stdv{4.17} & 86.14\stdv{2.03} & 86.03\stdv{2.00} &
89.52\stdv{2.93} & 42.30\stdv{2.64} & 42.21\stdv{1.80} &
43.85\stdv{2.18} & 48.77\stdv{3.11} \\
 & CGR & 81.43\stdv{4.17} & 85.90\stdv{2.06} & 86.03\stdv{2.00} &
89.52\stdv{2.93} & 42.30\stdv{2.64} & 42.21\stdv{1.80} &
43.85\stdv{2.18} & 48.77\stdv{3.11} \\
 & MGR & 81.43\stdv{4.17} & 86.14\stdv{2.03} & 86.03\stdv{2.00} &
89.52\stdv{2.93} & 42.30\stdv{2.64} & 42.21\stdv{1.80} &
43.85\stdv{2.18} & 48.77\stdv{3.11} \\
 & C-HMM & 79.43\stdv{4.28} & 85.54\stdv{2.08} & 85.08\stdv{2.12} &
89.52\stdv{2.93} & 43.03\stdv{2.64} & 42.21\stdv{1.79} &
43.95\stdv{2.18} & 47.95\stdv{3.09} \\
 & IJSR & 79.71\stdv{4.41} & 85.74\stdv{2.08} & 85.40\stdv{2.12} &
88.57\stdv{2.99} & 42.87\stdv{2.66} & 43.17\stdv{1.91} &
44.57\stdv{2.23} & 49.18\stdv{3.14} \\
 & AJSR & 85.43\stdv{3.86} & 86.49\stdv{2.08} & 86.67\stdv{2.06} &
89.52\stdv{2.93} & 42.62\stdv{2.66} & 42.67\stdv{1.83} &
44.26\stdv{2.20} & 50.00\stdv{3.18} \\
\rowcolor{cBest}
 & \method & \best{96.29}\stdv{0.95} & \best{90.66}\stdv{1.74} &
\best{91.75}\stdv{1.51} & \best{91.43}\stdv{2.67} &
\best{53.85}\stdv{2.51} & \best{52.45}\stdv{1.78} &
\best{54.00}\stdv{2.08} & \best{67.21}\stdv{3.07} \\
 & {\itshape $\Delta$ vs.\ best ctl.} &
\cellcolor{cPos!17}+10.86 & \cellcolor{cPos!8}+4.18 &
\cellcolor{cPos!8}+5.08 & \cellcolor{cPos!8}+1.90 &
\cellcolor{cPos!17}+10.82 & \cellcolor{cPos!15}+9.28 &
\cellcolor{cPos!15}+9.43 & \cellcolor{cPos!28}+17.21 \\
\midrule

Phi-4MM & DHC & 14.00\stdv{3.70} & 38.38\stdv{2.19} &
36.51\stdv{2.26} & 13.33\stdv{3.46} & 16.64\stdv{2.21} &
10.67\stdv{1.06} & 14.24\stdv{1.57} & 15.98\stdv{2.41} \\
 & ICI & 21.14\stdv{1.46} & 41.82\stdv{2.51} & 43.81\stdv{2.30} &
67.62\stdv{4.49} & 15.98\stdv{2.12} & 4.70\stdv{0.49} &
16.39\stdv{2.01} & 15.98\stdv{2.31} \\
 & CGR & 18.57\stdv{1.41} & 36.05\stdv{2.24} & 40.95\stdv{2.17} &
61.90\stdv{4.63} & 15.98\stdv{2.12} & 4.70\stdv{0.49} &
16.39\stdv{2.01} & 15.98\stdv{2.31} \\
 & MGR & 21.14\stdv{1.46} & 41.82\stdv{2.51} & 43.81\stdv{2.30} &
67.62\stdv{4.49} & 15.98\stdv{2.12} & 4.70\stdv{0.49} &
16.39\stdv{2.01} & 15.98\stdv{2.31} \\
 & C-HMM & 20.86\stdv{1.45} & 41.03\stdv{2.37} & 43.49\stdv{2.27} &
66.67\stdv{4.44} & 15.98\stdv{2.12} & 4.70\stdv{0.49} &
16.39\stdv{2.01} & 15.98\stdv{2.31} \\
 & IJSR & 20.86\stdv{1.49} & 41.65\stdv{2.53} & 43.49\stdv{2.37} &
66.67\stdv{4.56} & 15.98\stdv{2.12} & 4.70\stdv{0.49} &
16.39\stdv{2.01} & 15.98\stdv{2.31} \\
 & AJSR & 19.71\stdv{1.42} & 38.95\stdv{2.21} & 42.22\stdv{2.27} &
65.71\stdv{4.48} & 15.98\stdv{2.12} & 4.70\stdv{0.49} &
16.39\stdv{2.01} & 15.98\stdv{2.31} \\
\rowcolor{cBest}
 & \method & \best{90.86}\stdv{2.22} & \best{78.54}\stdv{2.64} &
\best{82.22}\stdv{2.25} & \best{80.00}\stdv{3.88} &
\best{18.93}\stdv{2.18} & \best{10.85}\stdv{1.04} &
\best{19.88}\stdv{1.99} & \best{20.90}\stdv{2.57} \\
 & {\itshape $\Delta$ vs.\ best ctl.} &
\cellcolor{cPos!80}+69.71 & \cellcolor{cPos!59}+36.72 &
\cellcolor{cPos!61}+38.41 & \cellcolor{cPos!20}+12.38 &
\cellcolor{cPos!8}+2.30 & \cellcolor{cPos!8}+0.18 &
\cellcolor{cPos!8}+3.48 & \cellcolor{cPos!8}+4.92 \\
\midrule

MiniCPM-o 2.6 & DHC & 30.00\stdv{5.54} & 42.42\stdv{2.34} &
42.86\stdv{2.22} & 33.33\stdv{4.62} & 18.69\stdv{2.19} &
14.42\stdv{1.30} & 20.39\stdv{2.06} & 20.08\stdv{2.52} \\
 & ICI & 79.71\stdv{4.12} & \best{84.67}\stdv{1.88} &
84.13\stdv{1.99} & \best{93.33}\stdv{2.38} & 20.08\stdv{2.20} &
13.75\stdv{1.07} & 22.23\stdv{2.08} & 22.13\stdv{2.68} \\
 & CGR & 79.71\stdv{4.12} & \best{84.67}\stdv{1.88} &
84.13\stdv{1.99} & \best{93.33}\stdv{2.38} & 20.08\stdv{2.20} &
13.74\stdv{1.07} & 22.23\stdv{2.08} & 22.13\stdv{2.68} \\
 & MGR & 79.43\stdv{4.21} & 84.16\stdv{1.91} & 83.81\stdv{2.03} &
92.38\stdv{2.53} & 19.92\stdv{2.20} & 13.51\stdv{1.07} &
22.03\stdv{2.08} & 21.72\stdv{2.65} \\
 & C-HMM & 77.71\stdv{4.32} & 83.72\stdv{1.98} &
83.17\stdv{2.11} & 92.38\stdv{2.53} & 20.16\stdv{2.19} &
13.94\stdv{1.08} & 22.34\stdv{2.08} & 22.13\stdv{2.68} \\
 & IJSR & 81.71\stdv{4.08} & 84.57\stdv{1.90} &
83.81\stdv{2.09} & 92.38\stdv{2.53} & 20.16\stdv{2.19} &
13.82\stdv{1.07} & 22.34\stdv{2.10} & 22.13\stdv{2.68} \\
 & AJSR & 79.14\stdv{4.24} & 83.87\stdv{1.98} &
83.49\stdv{2.11} & 92.38\stdv{2.53} & 19.92\stdv{2.22} &
13.50\stdv{1.16} & 22.03\stdv{2.12} & 20.90\stdv{2.57} \\
\rowcolor{cBest}
 & \method & \best{91.43}\stdv{2.83} & 82.95\stdv{2.27} &
\best{85.08}\stdv{1.97} & 91.43\stdv{2.64} &
\best{24.18}\stdv{1.97} & \best{22.61}\stdv{1.36} &
\best{25.20}\stdv{1.65} & \best{28.69}\stdv{2.85} \\
 & {\itshape $\Delta$ vs.\ best ctl.} &
\cellcolor{cPos!16}+9.71 & \cellcolor{cNeg!8}$-$1.72 &
\cellcolor{cPos!8}+0.95 & \cellcolor{cNeg!8}$-$1.90 &
\cellcolor{cPos!8}+4.02 & \cellcolor{cPos!13}+8.18 &
\cellcolor{cPos!8}+2.87 & \cellcolor{cPos!10}+6.56 \\
\bottomrule
\end{tabular}}
\end{table}

On HumDial-En, \method leads on S-BAcc and step accuracy for every model.
With Qwen2-Audio it gains $+40.57$ S-BAcc, $+21.59$ step accuracy, and
$+14.03$ step macro-F1 over the controlled envelope. With Phi-4MM it prevents
a collapse, raising S-BAcc by 69.71 points.

On CREMA-D-Stream, \method wins all four reported metrics on every backbone,
with its largest endpoint gain here ($+17.21$ final accuracy on
Qwen2.5-Omni). Across both benchmarks, \method is best on step accuracy and
S-BAcc in $8/8$ settings, on step macro-F1 in $7/8$, and on final accuracy in
$6/8$.

Appendix~\ref{app:published} additionally compares \method with the published
contextual model DVL-CER~\citep{zha2025dual}; \method is better in all 24
trajectory cells.

\subsection{\label{answer3}A4: Grounding Enables Effective Causal Revision}

Table~\ref{tab:ablation} separates grounding from filtering, applying the same
\emph{deployed} filter, including its asymmetric kernel and shrinkage, in both
policy arms. The policy-only column therefore applies an asymmetry-aware filter
to the contaminated observation. Replacing that observation with a prior-blind
acoustic posterior is the larger effect in all eight settings. Causal filtering
then performs history-aware revision on the grounded posterior and further
improves \sba in seven of eight settings. On HumDial-En with Qwen2-Audio, the
same filter contributes $+8.29$ \sba before grounding and $+16.86$ after it,
showing that temporal revision benefits from a cleaner observation.

% ---------- sections/ablation_table.tex ----------
% Auto-generated by gen_main_table.py -- do not edit by hand.
\begin{table}[t]
\caption{\textbf{Component ablation on state-balanced accuracy} (\%, point
estimates). All arms replay the same episodes.
``Policy'' applies the deployed evidence-shrunk \emph{asymmetric} filter
(Eq.~\ref{eq:shrinkage}) to the \emph{contaminated} seed observation;
``Ground'' applies chunkwise updates to the prior-blind observation; and
``Both'' is the full firewalled inference path. The two stages of this path are
complementary: grounding restores the current-audio observation, while causal
filtering incorporates temporal state only after perception has been decoupled
from historical belief.}
\label{tab:ablation}
\centering
\scriptsize
\setlength{\tabcolsep}{2.4pt}
\begin{tabular*}{\textwidth}{@{\extracolsep{\fill}}lrrrrrrrrrrrr}
\toprule
 & \multicolumn{6}{c}{HumDial-En} &
 \multicolumn{6}{c}{CREMA-D-Stream} \\
\cmidrule(lr){2-7}\cmidrule(l){8-13}
Model & Neither & Policy & Ground & Both & $\Delta$grd. & $\Delta$flt.
      & Neither & Policy & Ground & Both & $\Delta$grd. & $\Delta$flt. \\
\midrule
Qwen2-Audio & 29.14 & 37.43 & 53.71 & \best{70.57} &
\cellcolor{cPos!27}+24.57 & \cellcolor{cPos!19}+16.86 &
70.33 & 68.69 & \best{74.75} & 74.02 &
\cellcolor{cPos!8}+4.43 & \cellcolor{cNeg!8}$-$0.74\\
Qwen2.5-Omni & 81.43 & 86.29 & 91.43 & \best{96.29} &
\cellcolor{cPos!11}+10.00 & \cellcolor{cPos!8}+4.86 &
42.30 & 45.00 & 50.66 & \best{53.85} &
\cellcolor{cPos!9}+8.36 & \cellcolor{cPos!8}+3.20 \\
Phi-4MM & 21.14 & 28.57 & 84.29 & \best{90.86} &
\cellcolor{cPos!69}+63.14 & \cellcolor{cPos!8}+6.57 &
15.98 & 16.39 & 18.69 & \best{18.93} &
\cellcolor{cPos!8}+2.70 & \cellcolor{cPos!8}+0.25 \\
MiniCPM-o 2.6 & 79.71 & 80.57 & 90.29 & \best{91.43} &
\cellcolor{cPos!12}+10.57 & \cellcolor{cPos!8}+1.14 &
20.08 & 21.31 & 24.02 & \best{24.18} &
\cellcolor{cPos!8}+3.93 & \cellcolor{cPos!8}+0.16 \\
\bottomrule
\end{tabular*}
\end{table}

\begin{figure}[t]
\centering
\includegraphics[width=\textwidth]{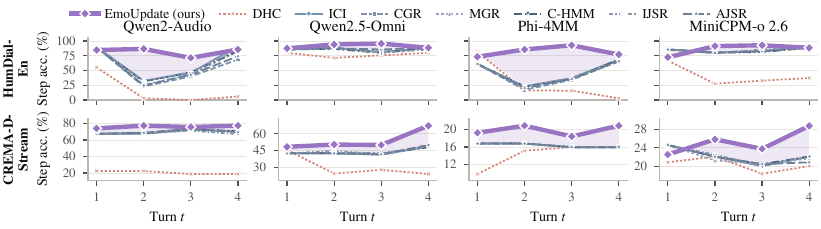}
\caption{\textbf{Causal step accuracy over the streaming trajectory}, all
backbones and controlled baselines. Shading is \method's margin over the best
baseline (leads / trails). The CREMA-D-Stream row uses \emph{per-panel}
$y$-ranges; read the ticks first.}
\label{fig:trajectory-curves}
\end{figure}

Figure~\ref{fig:trajectory-curves} shows where the gains arise:
seed-posterior policies collapse at the second turn on Qwen2-Audio and
Phi-4MM, the first turn at which a previous belief exists, while \method stays
nearly flat. Where the backbone is already strong, the curves separate less,
consistent with contamination rather than capacity being the limit.

\subsection{\label{answer5}A5: Robustness and Measurement-Parameterized Repairs}
\phantomsection\label{answer4}

\begin{figure}[t]
\centering
\includegraphics[width=\textwidth]{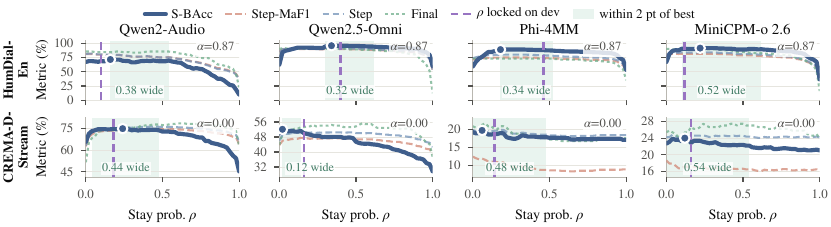}
\caption{\textbf{The locked stay-probability is not a delicate choice.}
Band: every $\rho$ within 2 S-BAcc points of the panel best (dot). Dashed line:
the $\rho$ locked on \emph{development} data, never read off this plot; inside
the band in six of eight settings and within 2.9 points of the optimum in all
eight. $\alpha$ is the coefficient of Equation~\ref{eq:shrinkage}.}
\label{fig:sensitivity}
\end{figure}

The four backbones react differently to history, yet the same design improves
step accuracy in every setting under an unchanged resource contract. The
51-point $\rho$ replay (Figure~\ref{fig:sensitivity}) shows a broad plateau
that every budget prefix (Table~\ref{tab:budget}) lands on. The shrinkage rule
adapts to the observed transition structure: HumDial-En development evidence
gives $\alpha=0.87$, lifting Qwen2-Audio's step accuracy from 69.8 to 82.2,
whereas CREMA-D-Stream, whose shifts are uniform by construction, returns
$\alpha=0.00$. In this case the transition prior reduces to the symmetric form
of Equation~\ref{eq:filter}, avoiding an unsupported asymmetric transition
structure.

Cross-validated on the balanced intervention grid, the operator of
Section~\ref{sec:decon} lifts exposed accuracy from 30.42 to 53.89
(control 72.50) and reduces wrong-prior pull from 71.00 to 12.17
(Appendix~\ref{app:decontamination}), without prompt access, training, or
additional SpeechLM calls.

\FloatBarrier
% ---------- sections/conclusion.tex ----------
\section{Conclusion}

We identified previous-belief contamination: a streaming SpeechLM's own
previous prediction can alter its perception of identical current audio, with
strong label asymmetry and the greatest risk where acoustic perception is
weakest. \method addresses this failure at its source by keeping current
perception prior-blind and introducing history only afterward through an
evidence-shrunk causal filter, with a training-free decontamination fallback
when firewalling is unavailable. Historical state should update a grounded
observation, never help create it.

\section*{Ethics Statement}
Emotion recognition is culturally and contextually uncertain, and systems
built on it should not be used for clinical diagnosis, hiring decisions, or
consequential surveillance. PBC adds a specific hazard: a streaming system can
amplify an erroneous earlier label into a confident trajectory. Deployments
should expose uncertainty and permit correction rather than treat filtered
beliefs as facts. All source corpora retain their original licenses, and
constructed trajectories introduce no generated emotion labels
(Appendix~\ref{app:reproducibility}).

\section*{Reproducibility Statement}
Every claim maps to a released artifact: episode manifests with data hashes,
perception caches, selection locks that freeze candidate order and chosen
parameters before evaluation, per-step predictions, bootstrap summaries,
plot-source tables, and a deterministic figure script.
Appendix~\ref{app:reproducibility} describes the six-stage locked reproduction
order; Appendices~\ref{app:data-construction} and~\ref{app:baselines} give
dataset construction, leakage controls, and exact baseline decision rules.
\label{sec:main_end}

\clearpage
\bibliography{references}
\bibliographystyle{iclr2027_conference}

\clearpage
\appendix
\begingroup
\parskip=2pt
{\small\tableofcontents}
\endgroup
\clearpage
% ---------- sections/appendix.tex ----------
\section{Extended Related Work}
\label{app:related}

\paragraph{Exposure bias and error accumulation.}
Sequence models trained with teacher forcing but run on their own predictions
accumulate error, the failure known as exposure bias.  Scheduled
sampling anneals ground-truth tokens toward model
samples~\citep{bengio2015scheduled}, sequence-level training optimizes the
evaluation metric directly~\citep{ranzato2015sequence}, and professor forcing
matches the free-running and teacher-forced dynamics
adversarially~\citep{lamb2016professor}.  Later analyses tie the effect to
hallucination and domain shift in translation~\citep{wang2020exposure}, question
how large it is for open-ended generation~\citep{schmidt2019generalization}, and
recast it as compounding error in imitation
learning~\citep{arora2022exposure}.  Every one of these remedies acts on
\emph{how the next token is produced} while the input representation is held
fixed.  PBC is the complementary failure: the input representation itself is
rewritten by the previous output, so the same audio yields a different
posterior.  This is why our repair is a firewall on the perception call rather
than a decoding rule, and why the diagnosis needs a counterfactual rather than a
held-out split.

\paragraph{Language models conditioned on asserted beliefs.}
Prompted LLMs are movable by what the context asserts.  Sycophancy makes
assistants agree with a stated user position~\citep{sharma2024towards}, an
effect that scales with model-written evaluation
pressure~\citep{perez2022discovering}, and persuasive conversation can overturn
a model's factual belief~\citep{xu2024earth}.  Stated reasoning need not be
the operative cause~\citep{turpin2023language}, and models do not reliably
repair their own errors when asked~\citep{huang2024large}; aggregating
independent samples helps precisely because a single conditioned trajectory is
unreliable~\citep{wang2022self}.  Purely
positional factors move predictions as well: majority-label and recency bias in
few-shot prompts~\citep{zhao2021calibrate}, example
ordering~\citep{lu2022fantastically}, and the position of relevant evidence in a long
context~\citep{liu2024lost}.  Our contaminating variable is narrower and
recurrent: a single label the system itself emitted one chunk earlier, re-entering
the prompt at every step.

\paragraph{Causal intervention as an analysis tool.}
Holding an input fixed and editing one internal or textual variable is the
standard way to move from correlation to mechanism in NLP analysis: causal mediation for attribute bias~\citep{vig2020investigating} and activation-level
localization of factual associations~\citep{meng2022locating}.  Our protocol applies
the same logic at the prompt surface of a frozen SpeechLM: audio, instruction,
and decoding are fixed and only the injected previous label varies.

\paragraph{Speech representations, corpora, and audio-language models.}
Self-supervised encoders provide the acoustic front end most emotion systems
build on~\citep{baevski2020wav2vec,hsu2021hubert,chen2022wavlm}.  Emotion corpora
span acted dyads~\citep{busso2008iemocap}, naturalistic
podcasts~\citep{lotfian2017building}, and multi-party
dialogue~\citep{poria2019meld}, with conversational emotion recognition surveyed
in~\citet{poria2019emotion}.  Instruction-following audio-language models
extend this to open-ended
listening~\citep{tang2024salmonn,gong2024listen,chu2024qwen2,xu2025qwen25omnitechnicalreport}.
We hold such a model frozen and treat its posterior as the observation of a
filter rather than as a final answer; the endpoint prior correction of
Section~\ref{sec:decon} is a label-shift adjustment, distinct from confidence
calibration in the temperature-scaling sense~\citep{guo2017calibration}.

\section{Metric Definitions}
\label{app:metrics}

Let $c_t=\mathbb{1}[y_t\neq y_{t-1}]$ and
$\hat c_t=\mathbb{1}[\hat y_t\neq\hat y_{t-1}]$.  A transition true positive
requires $c_t=1$, $\hat c_t=1$, and $\hat y_t=y_t$.  A predicted change in a
stable region is a false positive; a true shift without a correct landing is a
false negative.  TF1 is the resulting F1 score.  Change-state accuracy is
$\Pr(\hat y_t=y_t\mid c_t=1)$, stable-state accuracy is
$\Pr(\hat y_t=y_t\mid c_t=0)$, and \sba is their arithmetic mean.

Inertia is measured only when the previous prediction was correct: among steps
with $c_t=1$ and $\hat y_{t-1}=y_{t-1}$, it is the fraction for which the policy
does not change.  Volatility is also conditioned on a previously correct
prediction: among stable steps with $\hat y_{t-1}=y_{t-1}=y_t$, it is the
fraction for which $\hat y_t\neq y_t$.  This conditioning prevents already-wrong
states from being counted as evidence of either desirable stability or
destructive revision.

\section{Dataset Construction and Leakage Controls}
\label{app:data-construction}

\paragraph{CREMA-D-Stream.}
CREMA-D-Stream is constructed for this work from the original CREMA-D
corpus~\citep{cao2014crema}, a collection of acted utterances recorded from
professional actors over a fixed sentence set.  We reorganize its utterances
into four-chunk same-speaker streaming trajectories for causal trajectory
evaluation.  Each trajectory is assembled from four unique source utterances
and contains a single speaker.  A stable trajectory has the form
$(A,A,A,A)$, while a change trajectory has the form $(A,A,B,B)$ with
$A\neq B$.  Each development speaker contributes two trajectories of each
type; the evaluation split is constructed identically from disjoint speakers.
The resulting development and evaluation sets contain 120 and 244 episodes,
respectively.  Across both splits, 1,456 source clips are selected without
reuse.  The 30 development and 61 evaluation speakers have zero overlap.

\paragraph{HumDial-En.}
We preserve each Task-1 dialogue as an episode and expose the human-recorded
non-query turns in their original temporal order.  No future turn, reference
answer, emotion cause, or Task-4 conflict annotation is supplied to the model.
Development uses source group \texttt{0007}; evaluation uses groups
\texttt{0006}, \texttt{0008}, and \texttt{0010}.  This yields 45 development
and 105 evaluation episodes with no source-group overlap.  Task 4 remains
separate from the main trajectory evaluation and is used only as an
acoustic--semantic diagnostic.

\begin{table}[h]
\caption{Construction and leakage audit; each audio file appears in at most
one trajectory.}
\label{tab:data-audit}
\centering
\small
\begin{tabular}{lll}
\toprule
Property & CREMA-D-Stream & HumDial-En \\
\midrule
Episode source & Same-speaker clips & Original dialogue order \\
Temporal pattern & Stable or one shift & Naturally occurring shifts \\
Split key & Speaker identity & Released source group \\
Dev/eval episodes & 120 / 244 & 45 / 105 \\
Dev/eval chunks & 480 / 976 & 135 / 315 \\
Label ontology & 6 & 7 \\
Dev/eval gold shifts & 60 / 122 & 75 / 175 \\
Rev./Keep (dev) & 180 / 300 & 120 / 15 \\
Rev./Keep (eval.) & 366 / 610 & 280 / 35 \\
Cross-split key overlap & 0 & 0 \\
Source-audio reuse & None & None across source groups \\
Future information & Never exposed & Never exposed \\
\bottomrule
\end{tabular}
\end{table}

\section{Controlled Baseline Definitions}
\label{app:baselines}

Table~\ref{tab:baseline-definitions} gives the exact causal decision rules used
for the controlled envelope.  Except for DHC, which deliberately tests direct
history exposure, each rule operates on a cached current-chunk posterior.  A
threshold is chosen using development episodes only and is then locked before
evaluation.

\begin{table}[h]
\caption{Controlled baseline rules ($\tau$: development-selected scalar).}
\label{tab:baseline-definitions}
\centering
\footnotesize
\setlength{\tabcolsep}{4pt}
\begin{tabular}{@{}lll@{}}
\toprule
Acronym & Name & Causal update rule \\
\midrule
DHC & Direct history conditioning & Put $\hat y_{t-1}$ in the SpeechLM prompt \\
ICI & Independent chunk inference & $\hat y_t=\arg\max_y q_t(y)$ \\
CGR & Confidence-gated revision & Revise if $\max_y q_t(y)\geq\tau$ \\
MGR & Margin-gated revision & Revise if $q_t(y^*)-q_t(\hat y_{t-1})\geq\tau$ \\
C-HMM & Causal HMM & Symmetric Markov prediction followed by Bayes update \\
IJSR & Instantaneous JSD revision & Revise if $D_{\rm JS}(q_t,q_{t-1})\geq\tau$ \\
AJSR & Accumulated JSD revision & Accumulate confidence-weighted JSD to $\tau$ \\
\bottomrule
\end{tabular}
\end{table}

All seven systems are causal and use at most one frozen-SpeechLM call per
chunk.  CGR, MGR, C-HMM, IJSR, and AJSR receive the same ten-candidate scalar
budget.  DHC and ICI have no tunable scalar.  These controls are deliberately
simple because they isolate the two stages of the firewalled inference path; the separate
DVL-CER adaptation in Appendix~\ref{app:published} provides a comparison to a
recent published contextual model under its necessarily different training
contract.

\section{Selected Human-Readable Prompts}
\label{app:prompts}

The grounding prompt is chosen on development data only, under a rule fixed
before evaluation: candidates whose final accuracy falls more than one
percentage point below the seed prompt are rejected, and the remaining
candidates are ranked lexicographically by \sba, then \tfone, step accuracy,
and final accuracy, with grid order breaking exact ties.  The pool contains ten
candidates per setting.  The same ten candidate families are fixed across the
four backbones of each benchmark; development-time selection is performed
separately for each model--benchmark setting, without redesigning the candidate
pool.  This rule is distinct from the belief-policy protocol
of Section~\ref{sec:setup}, which additionally uses step macro-F1 as its first
tie-breaker; both are recorded in the released selection locks.
Table~\ref{tab:prompt-selection} lists the selected candidate for each setting.
The candidate pool always contains ten prompts.  The exact prompt text, hashes,
and development metrics are stored in the released selection locks.

\begin{table}[h]
\caption{Selected prior-blind prompt per setting.}
\label{tab:prompt-selection}
\centering
\small
\begin{tabular}{lll}
\toprule
Benchmark & Model & Selected candidate \\
\midrule
CREMA-D-Stream & Qwen2-Audio & \texttt{acoustic\_first\_no\_copy} \\
CREMA-D-Stream & Qwen2.5-Omni & \texttt{acoustic\_first\_original} \\
CREMA-D-Stream & Phi-4MM & \texttt{current\_audio\_contract} \\
CREMA-D-Stream & MiniCPM-o 2.6 & \texttt{prior\_deleted\_check\_short} \\
HumDial-En & Qwen2-Audio & \texttt{seed\_acoustic\_short} \\
HumDial-En & Qwen2.5-Omni & \texttt{acoustic\_over\_semantic\_short} \\
HumDial-En & Phi-4MM & \texttt{prior\_deleted\_check\_short} \\
HumDial-En & MiniCPM-o 2.6 & \texttt{full\_utterance\_short} \\
\bottomrule
\end{tabular}
\end{table}

For example, the selected Qwen2-Audio CREMA-D-Stream prompt is:
\begin{quote}\small
Listen only to the CURRENT speech audio. Do not copy any previous label or label
mentioned in history. Choose the emotion supported by audible pitch, energy,
rhythm, pace, vocal tension, and voice quality. Reply with exactly one lowercase
label: neutral, happy, sad, angry, fearful, or disgust.
\end{quote}
The history firewall is also enforced by the input wrapper: no previous label,
belief distribution, prior utterance, filename, or speaker identifier is passed
to the SpeechLM perception call.

Five additional selected prompt bodies are listed below. Each candidate follows
the same prior-blind perception contract, with the output ontology instantiated
for the target benchmark.

\paragraph{CREMA-D-Stream, Qwen2.5-Omni
(\texttt{acoustic\_first\_original}).}
Listen only to the CURRENT speech audio. Identify emotion from audible pitch
contour, loudness, tempo, rhythm, pauses, vocal tension, and voice quality. Do
not use sentence meaning, history, or a previous label. Reply with exactly one
lowercase label: neutral, happy, sad, angry, fearful, or disgust.
\paragraph{CREMA-D-Stream, Phi-4MM (\texttt{current\_audio\_contract}).}
The answer must depend only on the CURRENT speech audio. Previous beliefs,
filenames, and speaker IDs are not evidence. Judge vocal emotion from pitch,
loudness, tempo, pauses, rhythm, tension, and voice quality. Reply with exactly
one lowercase label: neutral, happy, sad, angry, fearful, or disgust.
\paragraph{HumDial-En, Qwen2-Audio (\texttt{seed\_acoustic\_short}).}
Listen only to the CURRENT attached speech audio with no history or previous
label. Judge how the sentence is spoken from pitch, energy, rhythm, pace, and
voice quality. Reply with exactly one lowercase HumDial-En label and nothing
else.
\paragraph{HumDial-En, Qwen2.5-Omni
(\texttt{acoustic\_over\_semantic\_short}).}
Listen to the CURRENT speech audio and classify how it is spoken. Prefer vocal
prosody over literal sentence meaning or any previous belief. Use pitch,
energy, timing, tension, and voice quality. Reply with exactly one lowercase
HumDial-En label.
\paragraph{HumDial-En, Phi-4MM
(\texttt{prior\_deleted\_check\_short}).}
Listen only to the CURRENT speech audio. Choose the label you would choose if
every previous-belief word were deleted from the prompt. Use pitch, loudness,
tempo, rhythm, pauses, vocal tension, and voice quality. Reply with exactly one
lowercase HumDial-En label.

Table~\ref{tab:prompt-inventory} lists the ten families the pool spans.

\begin{table}[h]
\caption{Ten prompt families in the fixed development pool.}
\label{tab:prompt-inventory}
\centering
\small
\begin{tabular}{ll}
\toprule
Candidate family & Controlled design intent \\
\midrule
\texttt{seed\_acoustic\_short} & Minimal current-audio instruction \\
\texttt{acoustic\_first\_original} & Enumerate prosodic evidence \\
\texttt{prior\_distractor} & Mark prior labels as distractors \\
\texttt{acoustic\_first\_no\_copy} & Explicitly forbid label copying \\
\texttt{current\_audio\_contract} & Restrict all evidence to current audio \\
\texttt{acoustic\_over\_semantic} & Prefer delivery under modality conflict \\
\texttt{label\_anchor\_guard} & State that ontology labels are not evidence \\
\texttt{full\_utterance} & Require whole-chunk vocal judgment \\
\texttt{prior\_deleted\_check} & Counterfactually delete prior words \\
\texttt{compact\_prototypes} & Supply short acoustic class prototypes \\
\bottomrule
\end{tabular}
\end{table}

\section{Difficulty-Stratified Analysis}
\label{app:ambiguity}

Table~\ref{tab:ambiguity-full} expands Table~\ref{tab:ambiguity} to all five
difficulty groups.  The strongest separation occurs on HumDial-En and on
Qwen2-Audio CREMA-D-Stream, where the grounded observation supplies enough acoustic
evidence for causal revision to act on.  On the lower-information CREMA-D-Stream
posteriors ($19.6\%$ and $24.6\%$ chunkwise accuracy) margin-based difficulty is
less discriminative and the policy differences are correspondingly small
($p\geq0.42$); the same regime explains the large correct-prior gains on the
nominally ``easy'' CREMA-D-Stream groups ($+42.86$ for Phi-4MM, $+23.53$ for
MiniCPM-o), since a high margin there marks confident errors rather than easy
audio, while the policy gap stays at most $1.28$ points.  Exact group sizes and
paired tests are reported for every group, including the two smaller
short-and-ambiguous subsets on HumDial-En ($N=45$ for Phi-4MM, $N=54$ for
MiniCPM-o).

% ---------- sections/ambiguity_full_table.tex ----------
% Auto-generated by gen_main_table.py -- do not edit by hand.
\begin{table}[h]
\caption{Complete difficulty-stratified analysis for the three backbones on
which the balanced injection was run.  Development-set medians define the
groups: \emph{short} is duration below the median, \emph{ambiguous} is
prior-blind posterior margin below the median, \emph{hard} is both at once,
and \emph{easy} is neither.  Left: prior intervention on the balanced
sample.  Right: all policies replayed on one shared posterior cache, with exact
paired McNemar $p$ against independent chunk inference.}
\label{tab:ambiguity-full}
\centering
\scriptsize
\setlength{\tabcolsep}{2.0pt}
\begin{tabular}{lllrrrrrrrrr}
\toprule
 & & & \multicolumn{4}{c}{Prior intervention} & \multicolumn{5}{c}{Shared-cache policy} \\
\cmidrule(lr){4-7}\cmidrule(l){8-12}
Data & Model & Chunks & $N$ & No prior & Correct & Pull & $N$ & ICI & \method & $\Delta$ & $p$ \\
\midrule
HumDial-En & Qwen2-Audio & all & 70 & 81.43 & 98.57 & 36.19 & 315 & 70.16 & 82.54 & +12.38 & $<$1e--4 \\
 &  & short & 32 & 71.88 & 100.00 & 41.15 & 124 & 68.55 & 85.48 & +16.94 & $<$1e--4 \\
 &  & ambiguous & 28 & 57.14 & 96.43 & 63.69 & 162 & 47.53 & 71.60 & +24.07 & $<$1e--4 \\
 &  & hard & 14 & 42.86 & 100.00 & 67.86 & 69 & 49.28 & 79.71 & +30.43 & $<$1e--4 \\
 &  & easy & 24 & 100.00 & 100.00 & 15.97 & 98 & 94.90 & 94.90 & +0.00 & 1.00 \\
\addlinespace[2pt]
 & Phi-4MM & all & 70 & 77.14 & 92.86 & 18.33 & 315 & 78.73 & 82.22 & +3.49 & 0.019 \\
 &  & short & 32 & 90.62 & 93.75 & 15.10 & 124 & 82.26 & 83.87 & +1.61 & 0.688 \\
 &  & ambiguous & 33 & 57.58 & 84.85 & 30.30 & 127 & 58.27 & 66.93 & +8.66 & 0.019 \\
 &  & hard & 12 & 83.33 & 83.33 & 29.17 & 45 & 64.44 & 68.89 & +4.44 & 0.688 \\
 &  & easy & 17 & 94.12 & 100.00 & 8.82 & 109 & 92.66 & 92.66 & +0.00 & 1.00 \\
\addlinespace[2pt]
 & MiniCPM-o 2.6 & all & 70 & 92.86 & 100.00 & 14.76 & 315 & 83.81 & 85.08 & +1.27 & 0.125 \\
 &  & short & 32 & 96.88 & 100.00 & 13.54 & 124 & 85.48 & 86.29 & +0.81 & 1.00 \\
 &  & ambiguous & 29 & 82.76 & 100.00 & 26.44 & 146 & 69.18 & 71.92 & +2.74 & 0.125 \\
 &  & hard & 11 & 90.91 & 100.00 & 25.76 & 54 & 70.37 & 72.22 & +1.85 & 1.00 \\
 &  & easy & 20 & 100.00 & 100.00 & 5.83 & 99 & 95.96 & 95.96 & +0.00 & 1.00 \\
\addlinespace[2pt]
\midrule
CREMA-D-Stream & Qwen2-Audio & all & 120 & 80.00 & 100.00 & 85.50 & 976 & 74.69 & 76.23 & +1.54 & 0.040 \\
 &  & short & 66 & 86.36 & 100.00 & 90.61 & 472 & 72.67 & 75.85 & +3.18 & 0.006 \\
 &  & ambiguous & 61 & 63.93 & 100.00 & 86.89 & 488 & 60.45 & 63.52 & +3.07 & 0.040 \\
 &  & hard & 30 & 73.33 & 100.00 & 93.33 & 230 & 55.22 & 61.74 & +6.52 & 0.006 \\
 &  & easy & 23 & 95.65 & 100.00 & 77.39 & 246 & 88.62 & 88.62 & +0.00 & 1.00 \\
\addlinespace[2pt]
 & Phi-4MM & all & 120 & 20.00 & 45.83 & 46.00 & 976 & 19.57 & 19.88 & +0.31 & 0.664 \\
 &  & short & 66 & 22.73 & 42.42 & 41.52 & 472 & 20.76 & 20.13 & -0.64 & 0.508 \\
 &  & ambiguous & 58 & 20.69 & 46.55 & 48.97 & 521 & 18.62 & 19.19 & +0.58 & 0.664 \\
 &  & hard & 32 & 12.50 & 40.62 & 45.00 & 241 & 20.75 & 19.50 & -1.24 & 0.508 \\
 &  & easy & 28 & 3.57 & 46.43 & 49.29 & 224 & 20.54 & 20.54 & +0.00 & 1.00 \\
\addlinespace[2pt]
 & MiniCPM-o 2.6 & all & 120 & 29.17 & 65.00 & 59.00 & 976 & 24.59 & 25.20 & +0.61 & 0.581 \\
 &  & short & 66 & 31.82 & 65.15 & 56.97 & 472 & 23.31 & 22.46 & -0.85 & 0.627 \\
 &  & ambiguous & 66 & 25.76 & 66.67 & 63.64 & 477 & 21.80 & 22.43 & +0.63 & 0.749 \\
 &  & hard & 29 & 27.59 & 62.07 & 63.45 & 208 & 22.60 & 20.67 & -1.92 & 0.424 \\
 &  & easy & 17 & 29.41 & 52.94 & 56.47 & 235 & 31.06 & 32.34 & +1.28 & 0.648 \\
\addlinespace[2pt]
\bottomrule
\end{tabular}
\end{table}

\paragraph{Stable versus shift steps.}
Table~\ref{tab:dynamics} decomposes the same shared-cache replay by whether the
true state changed at that step.  To distinguish effective revision from simply
revising more often, stable and true-shift steps are reported separately.  On
HumDial-En, shift accuracy improves on all three backbones, while stable
accuracy improves on two and is preserved on the third. Stable accuracy rises
by up to $14.29$ points, shift accuracy by up to $19.43$, and the false-switch
rate \emph{falls} by $2.86$--$11.43$ points.
On CREMA-D-Stream, where transitions are uniform by construction and the shrinkage estimator returns $\alpha=0$, the filter behaves conservatively
instead, raising stable accuracy in all three settings while leaving shift
accuracy unchanged to within one point on two of them.
Table~\ref{tab:dynamics} reports the per-setting false-switch rates.

% ---------- sections/dynamics_table.tex ----------
% Auto-generated by gen_main_table.py -- do not edit.
\begin{table}[t]
\caption{\textbf{Stable and shift steps improve together where the
trajectory carries information.}  Decomposition of the shared-cache replay by
whether the true state changed at that step (\%).  A policy that merely
revises more often would buy shift accuracy at the cost of stable accuracy and
a higher false-switch rate.  On HumDial-En (human recordings with many true shifts), \method improves
shift accuracy on all three backbones while improving stable accuracy on two
and preserving it on the third, with fewer spurious switches.  On CREMA-D-Stream, whose transitions are uniform by
construction and where the shrinkage estimator returns $\alpha=0$
(Eq.~\ref{eq:shrinkage}), it is conservative rather than aggressive: stable
accuracy rises in all three settings without a gain on shift steps.}
\label{tab:dynamics}
\centering
\small
\setlength{\tabcolsep}{3.8pt}
\begin{tabular}{lllrrrrr}
\toprule
Data & Model & Method & Stable $\uparrow$ & Shift $\uparrow$ &
False sw. $\downarrow$ & Delay $\downarrow$ & TF1 $\uparrow$ \\
\midrule
HumDial-En & Qwen2-Audio & ICI & 40.00 & 67.43 & 71.43 & 0.07 & 88.89 \\
 &  & C-HMM & 40.00 & 66.86 & 68.57 & 0.07 & 87.57 \\
 &  & \method & \ours{54.29} & \ours{86.86} & \ours{60.00} & \ours{0.04} & \ours{92.31} \\
\addlinespace[1.5pt]
 & Phi-4MM & ICI & 88.57 & 80.00 & 17.14 & 0.01 & 95.68 \\
 &  & C-HMM & 88.57 & 80.00 & 17.14 & 0.01 & 95.68 \\
 &  & \method & \ours{97.14} & \ours{84.57} & \ours{8.57} & \ours{0.01} & \ours{96.21} \\
\addlinespace[1.5pt]
 & MiniCPM-o 2.6 & ICI & 91.43 & 89.14 & 17.14 & 0.02 & 96.28 \\
 &  & C-HMM & 91.43 & 89.14 & 17.14 & 0.02 & 96.28 \\
 &  & \method & \ours{91.43} & \ours{91.43} & \ours{14.29} & \ours{0.01} & \ours{96.85} \\
\addlinespace[1.5pt]
CREMA-D-Stream & Qwen2-Audio & ICI & 74.92 & 74.59 & 24.26 & 0.13 & 57.52 \\
 &  & C-HMM & 78.36 & 69.67 & 16.07 & 0.14 & 63.35 \\
 &  & \method & \ours{78.36} & \ours{69.67} & \ours{16.07} & \ours{0.14} & \ours{63.35} \\
\addlinespace[1.5pt]
 & Phi-4MM & ICI & 20.16 & 17.21 & 31.64 & 0.32 & 22.03 \\
 &  & C-HMM & 20.66 & 17.21 & 20.49 & 0.25 & 17.71 \\
 &  & \method & \ours{20.66} & \ours{17.21} & \ours{20.49} & \ours{0.25} & \ours{17.71} \\
\addlinespace[1.5pt]
 & MiniCPM-o 2.6 & ICI & 25.90 & 22.13 & 41.48 & 0.27 & 23.53 \\
 &  & C-HMM & 26.07 & 19.67 & 31.48 & 0.29 & 17.97 \\
 &  & \method & \ours{27.05} & \ours{21.31} & \ours{62.30} & \ours{0.57} & \ours{19.10} \\
\bottomrule
\end{tabular}
\end{table}

\begin{figure}[h]
\centering
\includegraphics[width=\textwidth]{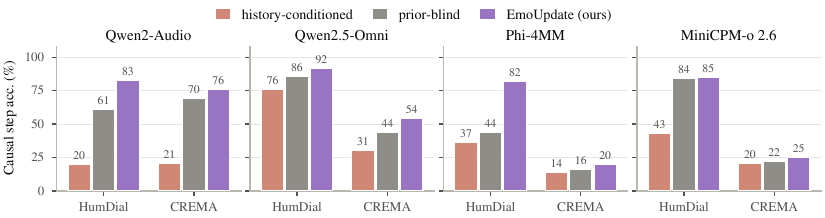}
\caption{\textbf{Contamination, and its repair, on every backbone.}  Writing
the previous label into the prompt (red) costs up to 49 points of causal step
accuracy relative to prior-blind inference (grey) across all eight settings;
\method (green) is highest in all eight.  Same quantities as the step-accuracy
columns of Table~\ref{tab:main-results}, drawn for comparison at a glance.}
\label{fig:contamination-all}
\end{figure}

\section{Compute and Fairness Audit}
\label{app:compute}

The test-time contract is identical in every row of Table~\ref{tab:compute}.
Prompt development calls are a one-time cost for choosing $s^\star$ and are not
repeated per policy or test seed.  More precisely, all belief policies share the
posterior cache produced by the selected grounding prompt.  We do not claim that
the unoptimized seed prompt and selected grounding prompt are the same.

The fairness claim is therefore specific: deployment calls, trainable
parameters, chunk boundaries, decoding, label extraction, and policy-candidate
budgets are matched.  Total development SpeechLM calls are not identical,
because prompt selection is a component of \method and costs the explicitly
reported ten prompt candidates.  Within any fixed perception source (seed or
selected grounding), every policy comparison replays exactly the same cache.

\begin{table}[!ht]
\caption{Compute audit: zero extra model calls, zero trainable parameters.}
\label{tab:compute}
\centering
\scriptsize
\setlength{\tabcolsep}{4pt}
\begin{tabular}{@{}llrrrrrr@{}}
\toprule
Data & Model & Test chunks & Calls/chunk & Test calls & Prompt cand. & Dev calls & Policy cand. \\
\midrule
\multirow{4}{*}{HumDial} & Qwen2-Audio & 315 & 1 & 315 & 10 & 1,350 & 10 \\
 & Qwen2.5-Omni & 315 & 1 & 315 & 10 & 1,350 & 10 \\
 & Phi-4MM & 315 & 1 & 315 & 10 & 1,350 & 10 \\
 & MiniCPM-o 2.6 & 315 & 1 & 315 & 10 & 1,350 & 10 \\
\addlinespace[2pt]
\multirow{4}{*}{CREMA-D-Stream} & Qwen2-Audio & 976 & 1 & 976 & 10 & 4,800 & 10 \\
 & Qwen2.5-Omni & 976 & 1 & 976 & 10 & 4,800 & 10 \\
 & Phi-4MM & 976 & 1 & 976 & 10 & 4,800 & 10 \\
 & MiniCPM-o 2.6 & 976 & 1 & 976 & 10 & 4,800 & 10 \\
\bottomrule
\end{tabular}
\end{table}

\section{Additional Intervention Results}
\label{app:intervention}

\begin{figure}[h]
\centering
\includegraphics[width=\textwidth]{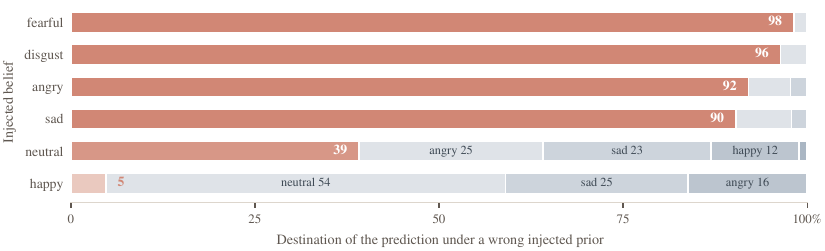}
\caption{\textbf{Where the prediction lands under each wrong injected prior}
(CREMA-D-Stream, Qwen2-Audio).  The leading terracotta segment is the share captured by the injected label (identical to Figure~\ref{fig:intervention}), and the slate segments give the destinations of the remainder.  Eligibility is the audit rule
used throughout: a condition counts only when the injected prior differs from
the prior-blind control prediction.  A wrong \emph{happy} prior is adopted
$4.8$\% of the time and its mass lands mostly on \emph{neutral}, whereas a
wrong \emph{fearful} prior is adopted $98.2$\% of the time.}
\label{fig:intervention-dest}
\end{figure}

Table~\ref{tab:intervention-full} reports all saved history-intervention
summaries.  CREMA-D-Stream uses six injected labels plus the control (seven calls per
sample).  HumDial uses seven injected labels plus the control (eight calls per
sample).  ``Evidence wins'' is the fraction of wrong-prior conditions where the
control was correct and the exposed prediction remains correct.  The Phi-4MM
accuracy increase under exposure is not evidence that history is harmless: its
43.67\% flip rate and posterior JSD of 0.30 show high sensitivity, while the
added label text acts as a regularizer for an otherwise weak control prompt.

\begin{figure}[h]
\centering
\includegraphics[width=0.66\textwidth]{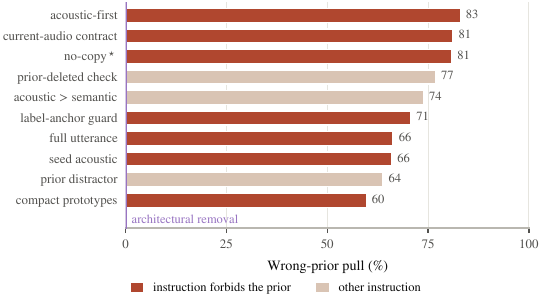}
\caption{All ten grounding candidates under the same balanced injection.  Dark
bars state explicitly that the previous label must not be used; $\star$ is the
deployed candidate, which never receives a prior at test time.  No instruction
falls below 59.7\% wrong-prior pull, the three most emphatic are the worst, and
pull rises with prior-blind accuracy ($r=0.81$).  The green rule marks what
architectural removal achieves.}
\label{fig:prompt-frontier}
\end{figure}

\begin{figure}[h]
\centering
\begin{minipage}[t]{0.495\textwidth}
  \centering
  \includegraphics[width=\linewidth]{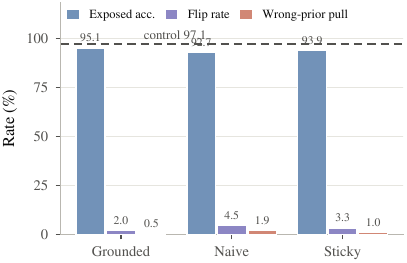}
  \centerline{\small (a) Qwen2.5-Omni}
\end{minipage}\hfill
\begin{minipage}[t]{0.495\textwidth}
  \centering
  \includegraphics[width=\linewidth]{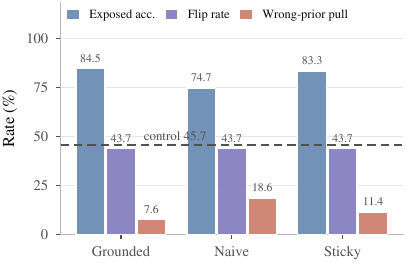}
  \centerline{\small (b) Phi-4MM}
\end{minipage}
\caption{HumDial-En previous-belief diagnostics under three prompt conditions.
Phi-4MM's accuracy rises under exposure while its flip rate holds at 43.67\%,
so accuracy change alone is not a contamination measure.}
\label{fig:humdial-conditions}
\end{figure}

Figure~\ref{fig:humdial-conditions} repeats the diagnostic on HumDial-En for
two further backbones.

\begin{table}[h]
\caption{Additional history-intervention summaries (\%); negative drop means
exposure helps.}
\label{tab:intervention-full}
\centering
\scriptsize
\setlength{\tabcolsep}{2.6pt}
\begin{tabular}{@{}lllrrrrrrrr@{}}
\toprule
Data & Model & Condition & $N$ & Ctrl. & Exposed & Drop & Flip & Wrong pull & Evidence wins & JSD \\
\midrule
\multirow{3}{*}{HumDial} & \multirow{3}{*}{Qwen2.5-Omni} & Grounded & 35 & 97.14 & 95.10 & 2.04 & 2.04 & 0.48 & 97.55 & 0.03 \\
        &          & Naive & 35 & 97.14 & 92.65 & 4.49 & 4.49 & 1.90 & 95.10 & 0.03 \\
        &          & Sticky & 35 & 97.14 & 93.88 & 3.27 & 3.27 & 0.95 & 97.06 & 0.02 \\
\addlinespace[2pt]
\multirow{3}{*}{HumDial} & \multirow{3}{*}{Phi-4MM} & Grounded & 35 & 45.71 & 84.49 & $-38.78$ & 43.67 & 7.62 & 98.96 & 0.30 \\
        &         & Naive & 35 & 45.71 & 74.69 & $-28.98$ & 43.67 & 18.57 & 96.88 & 0.29 \\
        &         & Sticky & 35 & 45.71 & 83.27 & $-37.55$ & 43.67 & 11.43 & 100.00 & 0.29 \\
\addlinespace[2pt]
CREMA-D-Stream & Qwen2-Audio & Prior-injected & 120 & 72.50 & 30.42 & 42.08 & 65.69 & 71.00 & 23.68 & 0.48 \\
\bottomrule
\end{tabular}
\end{table}

Table~\ref{tab:per-prior} resolves the balanced grid by injected label.

\begin{table}[h]
\caption{Per-prior breakdown, CREMA-D-Stream Qwen2-Audio (\%; JSD on 0--100).}
\label{tab:per-prior}
\centering
\small
\begin{tabular}{lrrr}
\toprule
Injected belief & Exposed accuracy $\uparrow$ & Prior pull $\downarrow$ & Posterior JSD $\downarrow$ \\
\midrule
neutral & 49.17 & 39.13 & 19.78 \\
happy & 50.00 & 4.76 & 23.24 \\
sad & 23.33 & 90.32 & 49.41 \\
angry & 21.67 & 92.05 & 50.61 \\
disgust & 20.00 & 96.40 & 62.68 \\
fearful & 18.33 & 98.20 & 83.67 \\
\bottomrule
\end{tabular}
\end{table}

\paragraph{Is the asymmetry a label-surface artefact?}
Pull is not explained by label spelling or by the model's own prior-blind base
rate.  Across
the six labels the pull rate is uncorrelated with surface length
($r=+0.020$) and with the prior-blind base rate
($r=+0.022$).  The clearest counterexample is internal:
\emph{happy} and \emph{angry} have identical spelling length and comparable base
rates (11.9\% versus 23.2\%)
yet differ by a factor of 19 in pull
(4.76\% versus 92.05\%); \emph{fearful}
has the \emph{lowest} base rate (6.4\%) and the
\emph{highest} pull.  The pattern is consistent with a semantic rather than a
tokenisation origin.

\section{Measurement-Based Decontamination}
\label{app:decontamination}

The intervention grid also supports repair when the perception prompt cannot
be firewalled, as in legacy or third-party serving stacks.  For every audited
sample the grid stores the clean posterior $q^{(\varnothing)}$ and the
contaminated posterior $q^{(b)}$ under each injected belief $b$.  We estimate the
per-prior log-offset $\delta_b$ of Equation~\ref{eq:decon-main} on one stratified
half of the grid and correct held-out contaminated posteriors as
$\hat q\propto q^{(b)}\exp(\delta_b)$.  The operator is closed-form, adds no
SpeechLM calls, and needs only the system's own previous label, which the deployed system itself wrote into the prompt.  Results are averaged over the two
folds.

\begin{figure}[h]
\centering
\includegraphics[width=0.52\textwidth]{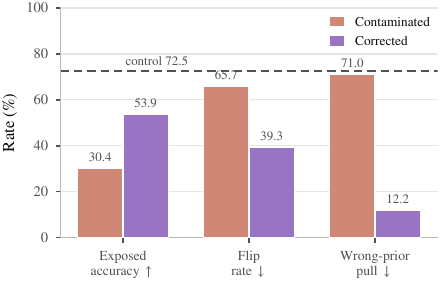}
\caption{Measurement-based decontamination (CREMA-D-Stream, Qwen2-Audio; two-fold
cross-validated).  The operator recovers 56\% of the accuracy damage and
collapses wrong-prior pull from 71.0 to 12.2.}
\label{fig:decontamination}
\end{figure}

Table~\ref{tab:decontamination} gives the per-fold numbers behind that figure.

\begin{table}[h]
\caption{Two-fold cross-validated decontamination, CREMA-D-Stream Qwen2-Audio
(\%).}
\label{tab:decontamination}
\centering
\small
\begin{tabular}{@{}lrrr@{}}
\toprule
Quantity & Contaminated & Corrected & Prior-blind control \\
\midrule
Exposed accuracy $\uparrow$ & 30.42 & 53.89 & 72.50 \\
Prediction flip rate $\downarrow$ & 65.69 & 39.31 & -- \\
Wrong-prior pull $\downarrow$ & 71.00 & 12.17 & -- \\
\bottomrule
\end{tabular}
\end{table}

The correction recovers 56\% of the accuracy damage
($30.42\to53.89$ against a $72.50$ control) and reduces the defining PBC
symptom, wrong-prior pull, from $71.00$ to $12.17$
(Figure~\ref{fig:decontamination}).

\paragraph{Deployment scope.}
The operator is intended as a fallback for serving stacks where firewalling is
unavailable, so the primary \method deployment uses firewalling whenever prompt
access is available.  The cross-validated analysis above evaluates the
correction on the CREMA-D-Stream/Qwen2-Audio intervention grid at the step level,
conditioned on the injected label, and requires only that the deployment-time prior is observed, which holds whenever the system itself maintains the belief
it exposes.

\paragraph{How much intervention grid does the operator need?}
We sub-sample the
training fold at fixed fractions, refit $\delta_b$, and evaluate on the
untouched held-out fold (three seeds $\times$ two folds).
Table~\ref{tab:decon-grid} shows the operator saturates almost immediately:
36 conditions (six chunks with all six priors injected) already recover exposed accuracy from
28.61 to
53.33, inside the seed-to-seed range of
the 360-condition fit
(53.89).  A practical deployment therefore
needs only a few dozen audited chunks, and refreshing is cheap.

\begin{table}[h]
\caption{Decontamination accuracy versus the size of the fitting grid
(CREMA-D-Stream, Qwen2-Audio; held-out fold, three seeds $\times$ two folds).
Uncorrected exposed accuracy is 28.61 and the
prior-blind control is 72.50.}
\label{tab:decon-grid}
\centering
\small
\begin{tabular}{rrrr}
\toprule
Fitting conditions & Exposed acc. $\uparrow$ & Range & Wrong-prior pull $\downarrow$ \\
\midrule
36 & 53.33 & 45.28--62.50 & 12.00 \\
72 & 54.54 & 47.50--58.06 & 14.44 \\
108 & 54.72 & 50.28--58.06 & 13.44 \\
144 & 54.44 & 51.11--57.22 & 13.89 \\
180 & 54.49 & 51.39--58.06 & 12.94 \\
216 & 54.54 & 50.83--57.22 & 12.33 \\
288 & 54.58 & 51.11--58.06 & 11.56 \\
360 & 53.89 & 52.78--55.00 & 12.17 \\
\bottomrule
\end{tabular}
\end{table}

\section{Asymmetry-Aware Transition Prior}
\label{app:asym}

Table~\ref{tab:asym} reports the full comparison behind
Section~\ref{answer5}.  Both variants replay identical posterior caches
with the same pre-specified ten-candidate $\rho$ grid; the asymmetric
leave distribution is an add-one-smoothed first-order estimate from
development gold sequences and introduces no additional tuned scalar.

% ---------- sections/shrinkage_table.tex ----------
% Auto-generated by gen_main_table.py -- do not edit by hand.
\begin{table}[h]
\caption{Transition-prior variants (\%, point estimates).  All variants
replay identical caches under the same ten-candidate budget.  $\alpha$ is the
shrinkage coefficient estimated from development transitions
(Equation~\ref{eq:shrinkage}): it is 0 on CREMA-D-Stream, where the evidence-shrunk
prior therefore coincides with the symmetric one by construction.}
\label{tab:asym}
\centering
\scriptsize
\setlength{\tabcolsep}{4pt}
\begin{tabular}{lllrrrr}
\toprule
Data & Model & Transition prior & Final & Step & TF1 & S-BAcc \\
\midrule
HumDial-En & Qwen2-Audio ($\alpha{=}0.87$) & Symmetric & \best{88.57} & 69.84 & 63.84 & 53.43 \\
 &  & Full asym. & 84.76 & 80.95 & 78.99 & \best{72.57} \\
 &  & Evidence-shrunk & 86.67 & 82.22 & 81.54 & 70.29 \\
\rowcolor{cBest}  &  & + endpoint calib. & 87.62 & \best{82.54} & \best{81.87} & 70.57 \\
\addlinespace[1.5pt]
 & Qwen2.5-Omni ($\alpha{=}0.87$) & Symmetric & \best{91.43} & 90.48 & 90.49 & 95.14 \\
 &  & Full asym. & 90.48 & 91.43 & 92.49 & 96.00 \\
 &  & Evidence-shrunk & 90.48 & 91.43 & 92.49 & 96.00 \\
\rowcolor{cBest}  &  & + endpoint calib. & \best{91.43} & \best{91.75} & \best{92.80} & \best{96.29} \\
\addlinespace[1.5pt]
 & Phi-4MM ($\alpha{=}0.87$) & Symmetric & 76.19 & 78.10 & 76.74 & 84.86 \\
 &  & Full asym. & 72.38 & 79.68 & 80.59 & 88.57 \\
 &  & Evidence-shrunk & 72.38 & 79.68 & 80.59 & 88.57 \\
\rowcolor{cBest}  &  & + endpoint calib. & \best{80.00} & \best{82.22} & \best{84.55} & \best{90.86} \\
\addlinespace[1.5pt]
 & MiniCPM-o 2.6 ($\alpha{=}0.87$) & Symmetric & \best{91.43} & 83.81 & 88.83 & 90.29 \\
 &  & Full asym. & 89.52 & 84.76 & 90.80 & 91.14 \\
 &  & Evidence-shrunk & 89.52 & 84.44 & 89.97 & 90.86 \\
\rowcolor{cBest}  &  & + endpoint calib. & \best{91.43} & \best{85.08} & \best{91.12} & \best{91.43} \\
\addlinespace[1.5pt]
\midrule
CREMA-D-Stream & Qwen2-Audio ($\alpha{=}0.00$) & Symmetric & \best{77.46} & \best{76.23} & \best{51.55} & 74.02 \\
 &  & Full asym. & 75.00 & 75.00 & 46.49 & \best{74.67} \\
 &  & Evidence-shrunk & \best{77.46} & \best{76.23} & \best{51.55} & 74.02 \\
\rowcolor{cBest}  &  & + endpoint calib. & \best{77.46} & \best{76.23} & \best{51.55} & 74.02 \\
\addlinespace[1.5pt]
 & Qwen2.5-Omni ($\alpha{=}0.00$) & Symmetric & 52.46 & 50.31 & \best{23.65} & 50.90 \\
 &  & Full asym. & 52.87 & 50.41 & 23.37 & 50.00 \\
 &  & Evidence-shrunk & 52.46 & 50.31 & \best{23.65} & 50.90 \\
\rowcolor{cBest}  &  & + endpoint calib. & \best{67.21} & \best{54.00} & 22.89 & \best{53.85} \\
\addlinespace[1.5pt]
 & Phi-4MM ($\alpha{=}0.00$) & Symmetric & \best{20.90} & \best{19.88} & 4.43 & \best{18.93} \\
 &  & Full asym. & 20.49 & \best{19.88} & \best{4.76} & \best{18.93} \\
 &  & Evidence-shrunk & \best{20.90} & \best{19.88} & 4.43 & \best{18.93} \\
\rowcolor{cBest}  &  & + endpoint calib. & \best{20.90} & \best{19.88} & 4.43 & \best{18.93} \\
\addlinespace[1.5pt]
 & MiniCPM-o 2.6 ($\alpha{=}0.00$) & Symmetric & 26.23 & 24.59 & \best{5.24} & 23.36 \\
 &  & Full asym. & 26.64 & 24.49 & 4.96 & 22.95 \\
 &  & Evidence-shrunk & 26.23 & 24.59 & \best{5.24} & 23.36 \\
\rowcolor{cBest}  &  & + endpoint calib. & \best{28.69} & \best{25.20} & 4.68 & \best{24.18} \\
\addlinespace[1.5pt]
\bottomrule
\end{tabular}
\end{table}

\paragraph{Finite-sample behaviour of $\alpha$.}
Equation~\ref{eq:shrinkage} is a point estimate, so we resample the development
destination transitions with replacement (2,000 draws) and recompute $\alpha$ on
each.  On HumDial-En ($n=75$ transitions) the coefficient is
$\alpha=0.87$ with a 95\% percentile interval
$[0.74,0.93]$, and \emph{no} resample returns zero.  On
CREMA-D-Stream ($n=60$) the point estimate is
$\alpha=0.00$ and 60\% of resamples
also return exactly zero, though the interval
$[0.00,0.64]$ is wide: with sixty transitions the data
cannot exclude moderate asymmetry, only fail to evidence it.  This is the intended behaviour: the estimator abstains rather than asserting structure, but it is an abstention, not a proof of symmetry.  Note $\alpha$ depends on the
benchmark's gold transitions alone, so it is shared by all four backbones.

\section{Conditional Volatility}
\label{app:full-results}

Table~\ref{tab:main-results} in the main paper reports the primary and
conventional metrics.  Table~\ref{tab:mechanism} adds the mechanism
diagnostics: transition F1, inertia, conditional volatility, and revision
delay.  The inertia--volatility pair exposes how the causal filter balances
persistence and revision: on HumDial-En \method reduces inertia to under
3.1\% for every backbone, while CREMA-D-Stream exhibits model-dependent trade-offs
between inertia and volatility.  Table~\ref{tab:volatility} reports conditional
volatility per method; it conditions on a previously correct stable belief and
therefore has empty denominators for some conservative policies.

% ---------- sections/mechanism_table.tex ----------
% Auto-generated by gen_main_table.py -- do not edit by hand.
\begin{table}[h]
\caption{Mechanism metrics (\%, point estimates; revision delay in chunks).
All four are lower-better except TF1.  Inertia is the fraction of true shifts
the policy fails to follow; volatility is the fraction of stable steps it
destroys; ``Best ctl.'' is the per-metric best controlled baseline.}
\label{tab:mechanism}
\centering
\small
\setlength{\tabcolsep}{5pt}
\begin{tabular}{lllrrrr}
\toprule
Data & Model & Method & TF1 $\uparrow$ & Inertia $\downarrow$ & Vol.\ $\downarrow$ & Delay $\downarrow$ \\
\midrule
HumDial-En & Qwen2-Audio & Best ctl. & 56.49 & 17.71 & 0.00 & 0.000 \\
\rowcolor{cBest} & & \method & 81.87 & 2.86 & 53.57 & 0.038 \\
\addlinespace[2pt]
 & Qwen2.5-Omni & Best ctl. & 87.68 & 3.47 & 9.52 & 0.013 \\
\rowcolor{cBest} & & \method & 92.80 & 2.50 & 0.00 & 0.012 \\
\addlinespace[2pt]
 & Phi-4MM & Best ctl. & 41.54 & 10.61 & 0.00 & 0.000 \\
\rowcolor{cBest} & & \method & 84.55 & 2.82 & 3.03 & 0.020 \\
\addlinespace[2pt]
 & MiniCPM-o 2.6 & Best ctl. & 85.29 & 7.30 & 9.09 & 0.000 \\
\rowcolor{cBest} & & \method & 91.12 & 2.88 & 9.09 & 0.012 \\
\midrule
CREMA-D-Stream & Qwen2-Audio & Best ctl. & 48.15 & 6.10 & 0.00 & 0.000 \\
\rowcolor{cBest} & & \method & 51.55 & 14.74 & 6.52 & 0.141 \\
\addlinespace[2pt]
 & Qwen2.5-Omni & Best ctl. & 18.36 & 15.09 & 25.77 & 0.306 \\
\rowcolor{cBest} & & \method & 22.89 & 6.15 & 24.83 & 0.367 \\
\addlinespace[2pt]
 & Phi-4MM & Best ctl. & 4.44 & 52.17 & 0.00 & 0.000 \\
\rowcolor{cBest} & & \method & 4.43 & 61.54 & 16.24 & 0.250 \\
\addlinespace[2pt]
 & MiniCPM-o 2.6 & Best ctl. & 4.84 & 75.00 & 0.83 & 0.091 \\
\rowcolor{cBest} & & \method & 4.68 & 56.25 & 53.47 & 0.567 \\
\bottomrule
\end{tabular}
\end{table}

% ---------- sections/volatility_table.tex ----------
% Auto-generated by gen_main_table.py -- do not edit by hand.
\begin{table}[h]
\caption{Conditional volatility (\%, lower is better; point estimates).
Dashes mark an empty conditional denominator, not a missing run.}
\label{tab:volatility}
\centering
\scriptsize
\setlength{\tabcolsep}{3.2pt}
\begin{tabular}{lrrrrrrrr}
\toprule
 & \multicolumn{4}{c}{HumDial-En} & \multicolumn{4}{c}{CREMA-D-Stream} \\
\cmidrule(lr){2-5}\cmidrule(l){6-9}
Method & Qwen2-A. & Qwen2.5 & Phi-4MM & MiniCPM & Qwen2-A. & Qwen2.5 & Phi-4MM & MiniCPM \\
\midrule
DHC & -- & 9.52 & 0.00 & 9.09 & 0.00 & 65.38 & 35.06 & 0.83 \\
ICI & 100.00 & 26.67 & 100.00 & 23.33 & 15.42 & 30.08 & 0.00 & 22.63 \\
CGR & -- & 26.67 & -- & 23.33 & 14.52 & 30.08 & 0.00 & 22.63 \\
MGR & -- & 26.67 & 100.00 & 23.33 & 13.82 & 30.08 & 0.00 & 22.46 \\
C-HMM & 50.00 & 30.00 & 100.00 & 24.14 & 13.58 & 30.89 & 0.00 & 23.19 \\
IJSR & 0.00 & 23.33 & 100.00 & 10.34 & 10.38 & 25.77 & 0.00 & 20.59 \\
AJSR & -- & 16.67 & -- & 20.69 & 10.44 & 28.02 & 0.00 & 18.98 \\
\rowcolor{cBest} \method & 53.57 & 0.00 & 3.03 & 9.09 & 6.52 & 24.83 & 16.24 & 53.47 \\
\bottomrule
\end{tabular}
\end{table}

\section{Published-Method Adaptation}
\label{app:published}

DVL-CER~\citep{zha2025dual} uses learned conversational representations rather
than frozen SpeechLM posteriors.  We therefore implement a causal posterior-GRU
adaptation following the published formulation and report it separately from the
compute-matched controlled envelope, since its randomized adaptation runs are
not the same uncertainty object as our episode-bootstrap replicates.
Table~\ref{tab:dvl} gives the complete comparison.

\begin{table}[h]
\caption{Published DVL-CER adaptation versus \method (\%).  Positive $\Delta$
favors \method; uncertainty estimates are reported in Table~\ref{tab:dvl}.}
\label{tab:published-main}
% NOTE: hand-maintained -- this summary is NOT produced by gen_main_table.py.
% Its \method columns must be kept in sync with Table~\ref{tab:dvl}
% (results/evidence_shrunk_hmm_v1/bootstrap_summary.md, evidence_shrunk_calibrated_endpoint).
\centering
\small
\setlength{\tabcolsep}{5.2pt}
\begin{tabular}{@{}llrrrrrr@{}}
\toprule
 & & \multicolumn{3}{c}{TF1 $\uparrow$} & \multicolumn{3}{c}{S-BAcc $\uparrow$} \\
\cmidrule(lr){3-5}\cmidrule(l){6-8}
Data & Model & DVL & \method & $\Delta$ & DVL & \method & $\Delta$ \\
\midrule
\multirow{4}{*}{CREMA-D-Stream} & Qwen2-Audio & 19.83 & \best{51.55} & \cellcolor{cPos!35}+31.72 & 48.25 & \best{74.02} & \cellcolor{cPos!28}+25.77 \\
 & Qwen2.5-Omni & 10.14 & \best{22.89} & \cellcolor{cPos!14}+12.75 & 32.05 & \best{53.85} & \cellcolor{cPos!24}+21.80 \\
 & Phi-4MM & 2.53 & \best{4.43} & \cellcolor{cPos!8}+1.90 & 17.10 & \best{18.93} & \cellcolor{cPos!8}+1.83 \\
 & MiniCPM-o 2.6 & 3.99 & \best{4.68} & \cellcolor{cPos!8}+0.69 & 19.33 & \best{24.18} & \cellcolor{cPos!8}+4.85 \\
\addlinespace[2.5pt]
\multirow{4}{*}{HumDial-En} & Qwen2-Audio & 13.23 & \best{81.87} & \cellcolor{cPos!76}+68.64 & 24.86 & \best{70.57} & \cellcolor{cPos!50}+45.71 \\
 & Qwen2.5-Omni & 18.01 & \best{92.80} & \cellcolor{cPos!80}+74.79 & 35.71 & \best{96.29} & \cellcolor{cPos!66}+60.58 \\
 & Phi-4MM & 16.53 & \best{84.55} & \cellcolor{cPos!75}+68.02 & 31.81 & \best{90.86} & \cellcolor{cPos!65}+59.05 \\
 & MiniCPM-o 2.6 & 23.53 & \best{91.12} & \cellcolor{cPos!75}+67.59 & 24.89 & \best{91.43} & \cellcolor{cPos!73}+66.54 \\
\bottomrule
\end{tabular}
\end{table}

% ---------- sections/dvl_table.tex ----------
% Auto-generated by gen_main_table.py -- do not edit by hand.
\begin{table}[h]
\caption{Paper-derived DVL-CER adaptation versus \method (\%).
DVL-CER reports mean$\pm$standard deviation over three randomized adaptation
runs; \method reports the full-set point estimate$\pm$episode-bootstrap
standard error ($B=1000$).}
\label{tab:dvl}
\centering
\scriptsize
\setlength{\tabcolsep}{4pt}
\begin{tabular}{lllrrr}
\toprule
Data & Model & Method & Step & TF1 & S-BAcc \\
\midrule
CREMA-D-Stream & Qwen2-Audio & DVL-CER adapt. & $51.57\pm6.33$ & $19.83\pm5.51$ & $48.25\pm5.29$ \\
\rowcolor{cBest}  &  & \method & $\best{76.23\pm2.06}$ & $\best{51.55\pm3.71}$ & $\best{74.02\pm2.58}$ \\
\addlinespace[1.5pt]
 & Qwen2.5-Omni & DVL-CER adapt. & $35.38\pm8.70$ & $10.14\pm4.40$ & $32.05\pm8.58$ \\
\rowcolor{cBest}  &  & \method & $\best{54.00\pm2.08}$ & $\best{22.89\pm2.44}$ & $\best{53.85\pm2.51}$ \\
\addlinespace[1.5pt]
 & Phi-4MM & DVL-CER adapt. & $17.49\pm0.90$ & $2.53\pm0.00$ & $17.10\pm0.67$ \\
\rowcolor{cBest}  &  & \method & $\best{19.88\pm1.99}$ & $\best{4.43\pm1.74}$ & $\best{18.93\pm2.18}$ \\
\addlinespace[1.5pt]
 & MiniCPM-o 2.6 & DVL-CER adapt. & $22.17\pm3.51$ & $3.99\pm2.31$ & $19.33\pm4.35$ \\
\rowcolor{cBest}  &  & \method & $\best{25.20\pm1.65}$ & $\best{4.68\pm1.29}$ & $\best{24.18\pm1.97}$ \\
\addlinespace[1.5pt]
\midrule
HumDial-En & Qwen2-Audio & DVL-CER adapt. & $21.38\pm8.89$ & $13.23\pm10.82$ & $24.86\pm31.70$ \\
\rowcolor{cBest}  &  & \method & $\best{82.54\pm2.02}$ & $\best{81.87\pm2.81}$ & $\best{70.57\pm4.16}$ \\
\addlinespace[1.5pt]
 & Qwen2.5-Omni & DVL-CER adapt. & $26.46\pm12.58$ & $18.01\pm16.72$ & $35.71\pm29.30$ \\
\rowcolor{cBest}  &  & \method & $\best{91.75\pm1.51}$ & $\best{92.80\pm1.86}$ & $\best{96.29\pm0.95}$ \\
\addlinespace[1.5pt]
 & Phi-4MM & DVL-CER adapt. & $24.23\pm12.86$ & $16.53\pm16.66$ & $31.81\pm30.65$ \\
\rowcolor{cBest}  &  & \method & $\best{82.22\pm2.25}$ & $\best{84.55\pm2.69}$ & $\best{90.86\pm2.22}$ \\
\addlinespace[1.5pt]
 & MiniCPM-o 2.6 & DVL-CER adapt. & $20.48\pm12.95$ & $23.53\pm12.71$ & $24.89\pm30.89$ \\
\rowcolor{cBest}  &  & \method & $\best{85.08\pm1.97}$ & $\best{91.12\pm2.04}$ & $\best{91.43\pm2.83}$ \\
\addlinespace[1.5pt]
\bottomrule
\end{tabular}
\end{table}

\section{Cross-Benchmark Component Analysis}
\label{app:components}

Table~\ref{tab:components-full} expands Table~\ref{tab:ablation} to every
reported metric, backbone, and benchmark, under the same bootstrap
protocol.

% ---------- sections/ablation_full_table.tex ----------
% Auto-generated by gen_main_table.py -- do not edit by hand.
\begin{table}[h]
\caption{Complete component analysis across four backbones and both benchmarks
(\%, point estimates).  ``Neither'': seed prompt with chunkwise updates;
``Policy only'': seed prompt with the causal filter; ``Ground only'':
prior-blind prompt with chunkwise updates; ``Both'': the deployed system.
Volatility is lower-better; a dash marks an empty conditional denominator.}
\label{tab:components-full}
\centering
\scriptsize
\setlength{\tabcolsep}{3.6pt}
\begin{tabular}{lllrrrrrr}
\toprule
Data & Model & Components & Final & Step & Step-MaF1 & TF1 & S-BAcc & Vol. \\
\midrule
HumDial-En & Qwen2-Audio & Neither & \best{89.52} & 60.95 & 68.28 & 50.75 & 29.14 & 100.00 \\
 &  & Policy only & \best{89.52} & 66.35 & 72.97 & 58.55 & 37.43 & 83.33 \\
 &  & Ground only & 87.62 & 70.16 & 71.40 & 63.33 & 53.71 & 61.54 \\
\rowcolor{cBest}  &  & Both (\method) & 87.62 & \best{82.54} & \best{82.31} & \best{81.87} & \best{70.57} & \best{53.57} \\
\addlinespace[1.5pt]
 & Qwen2.5-Omni & Neither & 89.52 & 86.03 & 86.14 & 86.61 & 81.43 & 26.67 \\
 &  & Policy only & 89.52 & 86.35 & 86.26 & 87.54 & 86.29 & 16.13 \\
 &  & Ground only & \best{91.43} & 90.16 & 89.55 & 90.86 & 91.43 & 9.38 \\
\rowcolor{cBest}  &  & Both (\method) & \best{91.43} & \best{91.75} & \best{90.66} & \best{92.80} & \best{96.29} & \best{0.00} \\
\addlinespace[1.5pt]
 & Phi-4MM & Neither & 67.62 & 43.81 & 41.82 & 35.80 & 21.14 & 100.00 \\
 &  & Policy only & 71.43 & 49.52 & 48.37 & 44.12 & 28.57 & 100.00 \\
 &  & Ground only & 76.19 & 78.73 & 74.88 & 77.81 & 84.29 & 6.45 \\
\rowcolor{cBest}  &  & Both (\method) & \best{80.00} & \best{82.22} & \best{78.54} & \best{84.55} & \best{90.86} & \best{3.03} \\
\addlinespace[1.5pt]
 & MiniCPM-o 2.6 & Neither & \best{93.33} & 84.13 & \best{84.67} & 84.57 & 79.71 & 23.33 \\
 &  & Policy only & \best{93.33} & 81.27 & 82.18 & 81.31 & 80.57 & 12.50 \\
 &  & Ground only & 91.43 & 83.81 & 82.10 & 88.83 & 90.29 & \best{6.45} \\
\rowcolor{cBest}  &  & Both (\method) & 91.43 & \best{85.08} & 82.95 & \best{91.12} & \best{91.43} & 9.09 \\
\addlinespace[1.5pt]
\midrule
CREMA-D-Stream & Qwen2-Audio & Neither & 70.08 & 69.57 & 68.41 & 39.80 & 70.33 & 15.42 \\
 &  & Policy only & 70.08 & 69.98 & 68.45 & 41.21 & 68.69 & 11.53 \\
 &  & Ground only & 74.59 & 74.69 & 73.44 & 45.91 & \best{74.75} & 13.94 \\
\rowcolor{cBest}  &  & Both (\method) & \best{77.46} & \best{76.23} & \best{74.58} & \best{51.55} & 74.02 & \best{6.52} \\
\addlinespace[1.5pt]
 & Qwen2.5-Omni & Neither & 48.77 & 43.85 & 42.21 & 17.65 & 42.30 & 30.08 \\
 &  & Policy only & 57.79 & 46.41 & 45.40 & 17.67 & 45.00 & 29.73 \\
 &  & Ground only & 52.87 & 50.41 & 47.23 & \best{23.61} & 50.66 & \best{24.75} \\
\rowcolor{cBest}  &  & Both (\method) & \best{67.21} & \best{54.00} & \best{52.45} & 22.89 & \best{53.85} & 24.83 \\
\addlinespace[1.5pt]
 & Phi-4MM & Neither & 15.98 & 16.39 & 4.70 & -- & 15.98 & \best{0.00} \\
 &  & Policy only & 18.03 & 16.91 & 9.17 & -- & 16.39 & 33.00 \\
 &  & Ground only & \best{20.90} & 19.57 & \best{11.20} & \best{4.52} & 18.69 & 29.06 \\
\rowcolor{cBest}  &  & Both (\method) & \best{20.90} & \best{19.88} & 10.85 & 4.43 & \best{18.93} & 16.24 \\
\addlinespace[1.5pt]
 & MiniCPM-o 2.6 & Neither & 22.13 & 22.23 & 13.75 & 4.12 & 20.08 & \best{22.63} \\
 &  & Policy only & \best{28.69} & 23.77 & 18.66 & 2.65 & 21.31 & 38.41 \\
 &  & Ground only & 25.41 & 24.59 & 17.18 & \best{6.12} & 24.02 & 33.79 \\
\rowcolor{cBest}  &  & Both (\method) & \best{28.69} & \best{25.20} & \best{22.61} & 4.68 & \best{24.18} & 53.47 \\
\addlinespace[1.5pt]
\bottomrule
\end{tabular}
\end{table}

The decomposition is consistent with the two-stage firewalled inference path: prior-blind
grounding supplies the largest change in observation quality, while the same
causal filter yields its strongest gains once that observation is usable.
Metric-specific trade-offs, including the CREMA-D-Stream settings where the filter
exchanges volatility or TF1 for balanced accuracy, are reported in
Table~\ref{tab:components-full}.

\section{Selection-Budget and Parameter Sensitivity}
\label{app:sensitivity}

Table~\ref{tab:budget} replays prefixes of the pre-specified joint grid.  Two
of the eight settings lock their final choice by three candidates; the others
keep refining up to the full budget.  The selection is nevertheless not
delicate: as Figure~\ref{fig:sensitivity} shows, the metric is flat across a
wide central band of $\rho$, so the candidates a smaller budget would have
chosen sit on the same plateau.

% ---------- sections/budget_table.tex ----------
% Auto-generated by gen_main_table.py -- do not edit by hand.
\begin{table}[h]
\caption{Selection-budget sensitivity.  Each cell is the candidate
($\rho$/endpoint mode) chosen when only the first $k$ of the ten pre-specified
joint candidates are available; ``def'' is the acoustic endpoint and ``LS'' the
label-shift endpoint.  A check mark means budgets 3, 5, and 10 agree.}
\label{tab:budget}
\centering
\small
\setlength{\tabcolsep}{6pt}
\begin{tabular}{llccccc}
\toprule
Data & Model & $k{=}1$ & $k{=}3$ & $k{=}5$ & $k{=}10$ & Stable \\
\midrule
HumDial-En & Qwen2-Audio & 0.1/def & 0.1/def & 0.1/def & 0.1/def & \checkmark \\
 & Qwen2.5-Omni & 0.1/def & 0.1/def & 0.4/def & 0.4/def &  \\
 & Phi-4MM & 0.1/def & 0.2/def & 0.4/def & 0.5/LS &  \\
 & MiniCPM-o 2.6 & 0.1/def & 0.1/LS & 0.1/LS & 0.1/LS & \checkmark \\
\midrule
CREMA-D-Stream & Qwen2-Audio & 0.1/def & 0.2/def & 0.4/def & 0.4/def &  \\
 & Qwen2.5-Omni & 0.1/def & 0.1/LS & 0.1/LS & 0.1/LS & \checkmark \\
 & Phi-4MM & 0.1/def & 0.2/def & 0.2/def & 0.2/def & \checkmark \\
 & MiniCPM-o 2.6 & 0.1/def & 0.1/LS & 0.1/LS & 0.1/LS & \checkmark \\
\bottomrule
\end{tabular}
\end{table}

That replay covers both benchmarks and all evaluated backbones.  Performance
degrades most visibly near $\rho=1$, where the filter approaches an
always-preserve policy and true revisions are suppressed; elsewhere the curves
are flat, which is why the budget prefixes in Table~\ref{tab:budget} land on
comparable operating points even when they select different candidates.

\clearpage
\section{Additional Diagnostic Analyses}
\label{app:failures}

Table~\ref{tab:negative-results} collects every pre-specified diagnostic case
in one place.

\begin{table}[h]
\caption{Pre-specified diagnostic cases across model--benchmark settings (gaps in points).}
\label{tab:negative-results}
\centering
\scriptsize
\setlength{\tabcolsep}{3.5pt}
\begin{tabular}{@{}lllp{4.40cm}@{}}
\toprule
Setting & Metric or diagnostic & Observation & Implication \\
\midrule
HumDial, Qwen2-Audio & Final accuracy & $-1.90$ vs.\ ICI & Final-only ranking hides trajectory gains \\
CREMA-D-Stream, Phi-4MM & TF1 & $-0.02$ vs.\ DHC & Low-information acoustic posterior \\
HumDial, Phi-4MM & Prior exposure & Accuracy $+38.78$, flip $43.67$ & Text may regularize yet still alter perception \\
HumDial stable steps & Volatility & Some denominators empty & Report ``--'' rather than impute a score \\
\bottomrule
\end{tabular}
\end{table}

The first row reinforces the need for trajectory metrics: \,\method improves
step accuracy, TF1, and S-BAcc by 21.59, 31.12, and 41.43 points against the
same baseline in the same setting.  The Phi-4MM CREMA-D-Stream row illustrates the
low-information regime: that current-audio posterior is heavily concentrated on
a small subset of labels, so temporal structure has limited acoustic evidence to
reweight.  The HumDial
intervention shows why accuracy change alone is not a PBC measure: sensitivity
is established by holding audio fixed and measuring output change, regardless
of whether the incidental change helps or hurts a particular label mix.

\section{Reproducibility and Responsible Use}
\label{app:reproducibility}

Every run stores episode manifests, posterior caches, selected-candidate locks,
configuration hashes, per-episode predictions, metric summaries, and plot-source
tables.  Table~\ref{tab:artifacts} maps each claim to its reproducibility
artifact.  Selection locks include candidate order, development metrics,
selected parameters, prompt hashes, data hashes, and cache hashes, preventing
post-evaluation candidate changes.

\begin{table}[h]
\caption{Reproducibility artifact inventory.}
\label{tab:artifacts}
\centering
\small
\begin{tabular}{ll}
\toprule
Artifact & Contents \\
\midrule
Episode manifests & Ordered chunks, labels, split keys, source-audio hashes \\
Perception caches & One dense label posterior per frozen-model chunk call \\
Selection locks & Candidate budget, development ranking, chosen prompt/policy \\
Policy predictions & Per-step prior, action, posterior, and predicted state \\
Bootstrap summaries & Point estimates with $B{=}1000$ bootstrap standard errors \\
Intervention records & Control and every injected-prior posterior \\
Plot-source CSVs & Every point and bar shown in the paper figures \\
Paper figure script & Deterministic vector/PDF figure generation \\
\bottomrule
\end{tabular}
\end{table}

\paragraph{End-to-end reproduction order.}
A clean reproduction follows six locked stages: (1) verify source-audio and
episode-manifest hashes; (2) run each frozen SpeechLM once per development
chunk for every prompt candidate; (3) select the prompt using only development
metrics and write the selection lock; (4) generate one evaluation posterior
cache with the selected prompt; (5) select the causal policy on cached
development posteriors and replay the locked policy on evaluation posteriors;
and (6) compute full-set metrics, $B{=}1000$ episode-bootstrap standard errors, and
figures from the saved predictions.  No evaluation label is read during prompt
or policy selection, and bootstrap seeds alter only episode weights, never model
outputs or selected hyperparameters.

The source data retain their original licenses; constructed trajectories
contain no generated emotion labels.  Emotion recognition is culturally and
contextually uncertain and should not be used as a clinical diagnosis, a hiring
signal, or a basis for consequential surveillance.  PBC also shows that system
state can amplify an erroneous earlier label; applications should expose uncertainty and
permit correction rather than treating filtered beliefs as facts.

\end{document}